\documentstyle[pra,aps,cite]{revtex}
\newcommand{\wa}{hydrogen molecule }
\newcommand{\bo}{Born-Oppenheimer approximation }
\newcommand{\co}{coordinate system }
\newcommand{\ha}{Hamiltonian }
\newcommand{\insmin}{\multicolumn{1}{c}{\hspace{0.3cm} \rule[1mm]{5mm}{0.4pt}}}
\newcommand{\zinsmin}{\multicolumn{1}{c}{\hspace{0.7cm} \rule[1mm]{7mm}{0.4pt}}}

\begin{document}
%\draft
\title{The hydrogen molecule in magnetic fields: 
The ground states of the $\Sigma$ manifold of the parallel configuration}
\author{T.Detmer, P. Schmelcher, F. K. Diakonos and L. S. Cederbaum}
\address{Theoretische Chemie, Physikalisch--Chemisches Institut,\\
Universit\"at Heidelberg, INF 253, D--69120 Heidelberg,\\
Federal Republic of Germany}
\maketitle

\begin{abstract}
The electronic structure of the \wa is investigated for the parallel configuration. 
The ground states of the $\Sigma$ manifold are studied for ungerade and gerade 
parity as well as singlet and triplet states covering a broad regime of field 
strengths from $B = 0$ up to $B = 100\;a.u.$. A variety of interesting phenomena can be observed. 
For the $^1\Sigma_g$ state we found a monotonous 
decrease of the equilibrium distance and a simultaneously increase of the dissociation energy 
with growing magnetic-field strength. The $^3\Sigma_g$ state is shown to develop an additional 
minimum which has no counterpart in field-free space. The $^1\Sigma_u$ state 
shows a monotonous increase in the dissociation energy with first 
increasing and than decreasing internuclear distance of the minimum. 
For this state the dissociation channel is $H_2 \rightarrow H^- + H^+$ 
for magnetic-field strengths $B \gtrsim 20\;a.u.$ due to the existence 
of strongly bound $H^-$ states in strong magnetic fields. The repulsive 
$^3\Sigma_u$ state possesses a very shallow van der Waals minimum for magnetic-field strengths 
smaller than $1.0\;a.u.$ within the numerical accuracy of our calculations. 
The $^1\Sigma_g$ and $^3\Sigma_u$ states cross as a function of B and 
the $^3\Sigma_u$ state, which is an unbound state, becomes the ground state of the 
\wa in magnetic fields $B \gtrsim 0.2\;a.u.$. 
This is of particular interest for the existence of molecular hydrogen 
in the vicinity of white dwarfs.  
In superstrong fields the ground state is again a strongly bound state, the $^3\Pi_u$ state. 

\end{abstract}


\section{Introduction}
During the past years the behavior and properties of particle systems in strong magnetic fields 
became a subject of increasing interest in different areas of physics like 
astrophysics or solid-state physics.
In astrophysics the discovery of strong magnetic fields in the vicinity of 
white dwarfs $\left(B \lesssim 10^5 T\right)$ and neutron stars 
$\left(B \lesssim 10^8 T\right)$ \cite{ostriker:1968,kemp:1970,truemper:1977} 
gave rise to many investigations. 
On the other hand, for highly excited atomic states the Coulomb forces and magnetic forces 
are of comparable orders of magnitude even for laboratory field strengths. 
In solid-state physics, excitons show strong field effects 
due to their small effective masses and large dielectricity constants \cite{chiu:1974}. 

A number of investigations were performed to reveal the structure and properties of small atoms 
and one-dimensional chains in such strong fields. They include variational methods 
\cite{glasser:1975,flowers:1977,mueller:1984}, Thomas-Fermi-type models 
\cite{fushiki:1992,abrahams:1992,lieb:1992,mueller:1971}, density-functional calculations 
\cite{jones:1985,jones:1986} and Hartree-Fock calculations \cite{virtamo:1976,
proeschel:1982,neuhauser:1987,miller:1991,lai:1992}. For a summary of energies, wave functions, 
and electromagnetic transitions concerning the hydrogen atom in strong 
magnetic fields, we refer the reader to Ref. \cite{ruder:1994}.

In contrast to the large number of studies on in particular the hydrogen atom 
in a strong magnetic fields there exist only a 
few investigations on molecular systems. Most of them deal with the $H_2^+ $ ion 
(see Refs. \cite{wille:1988,kappes1:1994,kappes2:1994,kappes:1995,kappes1:1996,kappes2:1996,kappes3:1996} 
and references therein) and little is known about the $H_2 $ molecule 
\cite{turbiner:1983,basile:1987,monteiro:1990,lai:1992,ortiz:1995,lai:1996}. 
In strong magnetic fields we encounter a variety of interesting new molecular phenomena. 
For the ground state of the $H_2^+ $ ion an increase of the 
electron density between the nuclear charges leads to a contraction of the bond length. 
At the same time we observe an increase in the dissociation energy with increasing 
magnetic-field strength. 
Moreover for the $H_2^+ $ ion a class of states with purely repulsive potential energy curves (PECs) in 
field-free space was shown to exhibit well-pronounced potential-energy minima 
in a sufficiently strong magnetic field \cite{kappes1:1994,kappes:1995}. 
Furthermore the topology of the electronic potential surfaces changes strongly 
with varying field strength. For intermediate field strengths it was shown, that 
the lowest-lying electronic states possess their global equilibrium configurations 
at positions corresponding to high symmetry, i.e., $\theta = 0\,^o$ or $\theta = 90\,^o$.  
However, for some excited states a 
global symmetry lowering occurs leading to global equilibrium configurations at $0\,^o < \theta < 90\,^o$ 
\cite{kappes1:1996,kappes2:1996,kappes3:1996}. 

Only little is known concerning the electronic 
structure of the \wa in a strong magnetic field. Most of the investigations deal with 
the \wa in superstrong fields $\left(B \ge 10^{11} G \right)$. 
For intermediate field strengths there exist only two studies of qualitative character 
which investigate the PEC of the lowest $^1\Sigma_g$ state \cite{basile:1987,turbiner:1983}. 
A detailed knowledge of the electronic structure of the \wa is of particular relevance 
in astrophysics since it might lead to a better 
understanding of the spectra of white dwarfs and neutron stars. 
Hereby the ground state of the \wa is of particular interest. 
Recently a controversial discussion arose concerning the electronic structure of a \wa 
in a superstrong magnetic field \cite{korolev:1992,korolev1:1993,korolev2:1993, 
korolev:1994,ortiz:1995,korolev:1995,lai:1995}. 
It was conjectured that in superstrong magnetic fields  
the $^3\Sigma_u $ state would be the ground state of $H_2$ \cite{korolev:1992}. 
This state possesses a very shallow van der Waals minimum in field-free space at $R \sim 8 a.u. $. 
Due to the spin-Zeeman shift the $^3\Sigma_u $ state monotonously decreases 
in energy with increasing magnetic-field strength. Therefore a crossing exists between 
the $^1\Sigma_g $ and the $^3\Sigma_u $ state at some magnetic-field strength $B_c $. 
For that reason the authors expected the weakly bound $^3\Sigma_u $ state to be the ground state 
of the \wa in superstrong fields. For such magnetic-field strengths, hydrogen might then 
be able to form a Bose-Einstein condensate and become superfluid. 
However it has been proved that the $^3\Pi_u $ state 
is the true ground state for magnetic-field strengths $B \gtrsim 3\times10^3\;a.u.$ 
\cite{ortiz:1995}. For magnetic-field strengths smaller than 
$B = 3\times 10^3\;a.u.$ the ground state of the \wa is not known. 

From the above it is evident that accurate investigations of the electronic structure of the \wa 
are very desirable. In the present investigation we perform a first step 
to elucidate the electronic properties of the \wa in a strong magnetic-field. 
Particular emphasis is placed on the intermediate regime which is of relevance to the 
physics of white dwarfs. We 
investigate the electronic structure of the lowest states of the $\Sigma$ manifold, i.e., 
the lowest singlet and triplet states with gerade and ungerade parity for a 
magnetic quantum number equal to zero. We hereby focus on the case of the parallel 
internuclear and magnetic-field axes. This configuration is distinct by its 
higher symmetry compared to the case of an arbitrary angle $\theta$ between the internuclear and 
magnetic-field axis. It has been shown \cite{schmelcher:1990} that the diabatic energy 
curves exhibit extrema at $\theta = 0^o$, and it can therefore be expected 
that the parallel configuration plays an important role. 

Our calculations will provide detailed data of equilibrium distances, 
total energies, and dissociation energies for the corresponding electronic states. 
Due to the efficiency of our method we are able to investigate the electronic structure 
in the complete range of field strengths, from $B = 0$ up to $B = 100\;a.u.$.

In detail the paper is organized as follows. In Sec. II we develop the theoretical framework 
as well as technical background for our calculations. Section III. is the central part of the paper. 
It contains the detailed discussion of the electronic PECs for the lowest 
$^1\Sigma_g,\;^1\Sigma_u,\;^3\Sigma_g$ and $^3\Sigma_u$ states for magnetic fields 
$B = 0\;-\;100\;a.u.$. The summary and conclusions are given in Sec. IV. 
We use a basis set of nonspherical atomic orbitals for our configuration-interaction calculations 
\cite{schmelcher3:1988}. Some details of this basis set are given in 
Appendix A. Furthermore, in Appendix A we describe the construction of the molecular wave function 
for the \wa. Appendix B contains a more detailed description of the computational techniques. 
The evaluation of the matrix elements and some special features of the implementation of our 
source code are discussed to some extent.

\section{Theoretical aspects}
We start with the total nonrelativistic molecular \ha in Cartesian coordinates.
It is well known that the total pseudomomentum is a constant of motion
\cite{johnson:1983,avron:1978}. For a neutral system like the \wa, 
the components of the pseudomomentum additionally commute with each other.
Therefore, the \ha can be simplified by performing a
so-called pseudoseparation of the center-of-mass motion 
\cite{schmelcher2:1988,schmelcher:1994,johnson:1983}. 
Due to this pseudoseparation the center-of-mass coordinate and the conserved 
pseudomomentum are introduced as a pair of canonical conjugated variables.
As a result, the center-of-mass coordinate does not appear in the transformed \ha.
After applying this transformation, the exact \ha can be further simplified 
by a series of unitary transformations.
For details of these transformations, we refer the reader to the literature
\cite{schmelcher2:1988,schmelcher:1994}.

In order to separate the electronic and nuclear motion an adiabatic approximation 
has to be performed, which means that we have to apply the \bo in the presence of a magnetic field.
The validity of the \bo in the presence of a magnetic field 
has been studied in detail by Schmelcher and coworkers
\cite{schmelcher1:1988,schmelcher2:1988,schmelcher:1994}, 
including all corrections due to finite nuclear mass.
In a first order approximation we choose the electronic \ha as the fixed nuclei
\ha, i.e., we assume infinitely heavy masses for the nuclei.
The origin of our \co coincides with the midpoint of the internuclear axis of the \wa
and the protons are located on the $z$ axis.  
The magnetic field is chosen parallel to the $z$ axis of our \co and
we use the symmetric gauge for the vector potential.
Finally our electronic \ha takes on the following appearance:
\begin{equation}
H =\sum\limits_{i=1}^2 \left\{\frac{1}{2}\bbox{p}_i^2 
+ \frac{1}{8}\left( \bbox{B}\times \bbox{r}_i\right)^2 
+ \frac{1}{2}\bbox{L}_i\hspace{-1pt}\cdot\hspace{-1pt}\bbox{B}- \frac{1}{|\bbox{r}_i - \bbox{R}/2|} 
- \frac{1}{|\bbox{r}_i + \bbox{R}/2|} \right\} \\
+ \frac{1}{|\bbox{r}_1 - \bbox{r}_2|} + \frac{1}{R} 
+ \bbox{S}\hspace{-1pt}\cdot\hspace{-1pt}\bbox{B}\label{form1}
\end{equation}
We hereby neclect relativistic effects such as, e.g., the spin-orbit coupling ,
and the gyromagnetic factor of the electron was chosen to be 2.
The symbols $\bbox{r}_i$, $\bbox{p}_i$, and $\bbox{L}_i$ denote the position vectors, the 
canonical conjugated momenta and the angular momenta of the two electrons, respectively.
$\bbox{B}$ and $\bbox{R}$ are the vectors of the magnetic field and internuclear distance, 
respectively and $R$ denotes the magnitude of $\bbox{R}$.
With $\bbox{S}$ we denote the vector of the total electronic spin.
Since we deal with $\Sigma$ states, the sum 
$\sum\limits_{i=1}^2 \bbox{L}_i\hspace{-1pt}\cdot\hspace{-1pt}\bbox{B}$ equals to zero. 
Throughout the paper we will use atomic units.

The \ha (\ref{form1}) commutes with the following operators:
 
a) The parity operator $P$ due to the charge symmetry of the molecule . 
The corresponding eigenfunctions are marked with the subscript g for gerade or u for ungerade
parity.
 
b) The projection $L_z$ of the electronic angular momentum on the internuclear axis. 

c) The square $S^2$ of the total electronic spin.
The electronic functions are labeled with a left superscript describing the 
multiplicity $2S +1$ of the state.

d) The projection $S_z$ of the total electronic spin on the internuclear axis.
For singlet states the only possibility is $M_s = 0$. For triplet states we are
only interested in the lowest state, which means $M_s = -1$.

In the absence of a magnetic field we encounter an additional symmetry namely the reflections 
with respect to the electronic coordinates at the $xz$ ($\sigma_v$) plane. 
The eigenfunctions possess the corresponding eigenvalues $\pm1$.
This symmetry does not hold in the presence of a magnetic field.
Therefore, the resulting symmetry groups for the \wa are $D_{\infty h}$ in the case of
field-free space, and $C_{\infty h}$ in the presence of a magnetic field \cite{schmelcher:1990}.

In order to solve the fixed-nuclei electronic Schr\"odinger equation 
belonging to the \ha (\ref{form1}), we expand the electronic 
eigenfunctions in terms of molecular configurations.
First of all we note that the total electronic wave function $\Psi_{ges}$ can 
be written as a product of its spatial part $\Psi$ and its spin part 
$\chi$, i.e., we have $\Psi_{ges} = \Psi \chi$.
For the spatial part $\Psi$ of the wave function we use the 
linear combination of atomic orbitals molecular orbital ansatz (LCAO-MO), i.e.,
we decompose $\Psi$ with respect to molecular orbital configurations $\psi$ of $H_2$ 
which respect the corresponding symmetries (see above) and the Pauli principle, i.e 
\begin{eqnarray}
\Psi &=& \sum\limits_{i,j} c_{ij} \left[\psi_{ij}\left(\bbox{r}_1,\bbox{r}_2\right) \pm 
\psi_{ij}\left(\bbox{r}_2,\bbox{r}_1\right)\right] \nonumber \\
&=& 
\sum\limits_{i,j} c_{ij} \left[\Phi_i\left(\bbox{r_1}\right)\Phi_j\left(\bbox{r_2}\right) 
\pm \Phi_i\left(\bbox{r_2}\right)\Phi_j\left(\bbox{r_1}\right)\right] \nonumber
\end{eqnarray}
The molecular-orbital configurations $\psi_{ij}$ of $H_2$ are products of the corresponding 
one-electron $H_{2}^+$ molecular orbitals $\Phi_i$ and $\Phi_j$.
The $H_{2}^+$ molecular orbitals are built from atomic orbitals 
centered at each nucleus. A key ingredient of this procedure is a
basis set of nonorthogonal optimized 
nonspherical Gaussian atomic orbitals which has been established previously 
\cite{schmelcher3:1988,kappes2:1994}.
A more detailed description of the construction of the molecular electronic wave function 
is presented in Appendix A.
In order to determine the molecular electronic wave function of $H_2$ 
we use the variation principle. This means we minimize the variational integral
$\frac{\int \Psi^* H \Psi}{\int \Psi^* \Psi}$ by varying the 
coefficients $c_i$.
The resulting generalized eigenvalue problem reads as follows:
\begin{equation}
\left(\underline{H} - \epsilon\underline{S}\right)\bbox{c} = \bbox{0} \label{form3}
\end{equation}
In the present investigation for parallel internuclear axis and magnetic-field axis, 
the \ha matrix $\underline{H}$ is real and symmetric and the overlap matrix
is real, symmetric and positiv definite. The vector $\bbox{c}$ contains the expansion coefficients.
The matrix elements of the \ha matrix and the overlap matrix
are certain linear combinations of matrix elements with respect to 
the optimized nonspherical Gaussian atomic orbitals.
The latter matrix elements were already calculated in Ref. \cite{schmelcher3:1988}.
However, the formulas for the electron-nucleus and in particular 
the electron-electron matrix elements given in Ref.\cite{schmelcher3:1988}
turned out to be not sufficiently efficient for numerical calculations with large basis sets.
Both, the numerical stability and the efficient and fast computation of 
the matrix elements required a different approach to the integral evaluation 
within our basis set of atomic orbitals. 
In our computational scheme the matrix elements are evaluated by a combination of a 
special quadrature method and a subsequent numerical integration. 
We herefore wrote a numerical integration routine which is very carefully 
adjusted to the special problem of the \wa parallel to a magnetic field.
We were able to reduce the mean time of one integral calculation 
by a factor of 1000. 
Appendix B deals with the most important aspects of the corresponding computational method. 
For the construction of the Hamilton matrix within each subspace we used a direct 
method which means that each integral is evaluated only once. 
This procedure is very useful since the same atomic orbitals contribute 
to different molecular orbitals. 
In addition, all four matrices belonging to different subspaces 
(parity, spin multiplicity) of the $\Sigma$ states 
are simultaneously constructed since they depend on the same matrix elements 
of the corresponding atomic orbitals. 

For each PEC about 300 points were calculated on an average. 
In the present calculations, the typical dimension of the \ha matrix for each subspace varies between
approximately 800 and 3000 depending on the magnetic-field strength.  
Depending on the dimension of the Hamiltonian matrix, it takes between 30 and 250 minutes for
simultaneously calculating one point of a PEC of each $\Sigma-$ subspace on a AIX 590 computer. 
Therefore, the total amount of CPU time for our calculations amounts to 
approximately seven months on the above computer. 
Previous to the calculation of the PECs we performed 
detailed test on the accuracy and convergence of the calculations. 
For each magnetic-field strength we developed a basis set of atomic orbitals 
particularly adapted for minimizing the total energy. 
The overall accuracy of our results is estimated to be better than $10^{-3}$.
For the numerical solution of the eigenvalue problem (\ref{form3}) we
used the standard NAG library.

\section{Results and discussion}
\subsection{The $^1\Sigma_g$ state}
Let us begin our investigation by considering the lowest $^1\Sigma_g$ state of the \wa,  
which is the ground state in field-free space.
This state has been extensively studied in field-free space in the adiabatic approximation, 
both theoretically and experimentally. 
For the literature on theoretical investigations up to 1960 
we refer the reader to the bibliography in Ref. \cite{mclean:1960}. 
In particular we mention the pioneering work of James and Coolidge \cite{james:1933}.  
They represent the wave function in elliptic coordinates and demonstrated  
the usefulnes of explicitely including the interelectronic distance in the wave function. 
A well-known drawback of the expansion in elliptical coordinates is that it 
converges more and more slowly with increasing values of the internuclear distance. 
In addition, the results of earlier work do not possess the accuracy needed for 
comparison with experimental data. Recently, several theoretical investigations were 
performed to improve the overall energy values for the $^1\Sigma_g^+$ state. 
Kolos \cite{kolos:1994} as well as Wolniewicz \cite{wolniewicz:1995} improved the 
electronic energy calculated in the \bo several times. As a reference in the case of 
field-free space, we use the energy values obtained by Wolniewicz in 1995 \cite{wolniewicz:1995}.

In field-free space the energy curve of the $^1\Sigma_g^+$ state shows only one minimum at the 
equilibrium distance of $1.4\;a.u.$ with a total energy of $-1.1744757\; a.u.$. In the dissociation limit 
we have two H atoms in their ground states, i.e., 
$H_2 \rightarrow H\left(1s\right) + H\left(1s\right)$. 
Therefore, in the separated atom limit the total energy approaches $-1.0\; a.u.$ 
which corresponds to the energy of two H atoms in the ground state. 
In the united atom limit we have a helium atom in the $^1S \; 1s^2$ state. 
For that reason the total energy without the nucleus-nucleus repulsion $\frac{1}{R}$ 
approaches an energy value of $2.90372\; a.u.$ with decreasing internuclear distance 
which is the ground state energy of the helium atom.

For the total energy at the equilibrium distance in field-free space 
we obtained a total energy of $-1.173892\; a.u.$ which yields 
a dissociation energy of $0.173892 \; a.u.$. This corresponds to a relative accuracy 
in the total energy of 
about $6\times 10^{-4}$ compared to the bench mark result in Ref. \cite{wolniewicz:1995}. 
This accuracy in the total energy even increases with increasing internuclear distance. For 
$R = 3 \;a.u.$ and $R  =  4\; a.u.$ we yield an error of 
$3\times10^{-4}$ and $2\times10^{-4}$, respectively. 

Let us explain how we obtained such an accuracy for a configuration-interaction (CI) calculation 
with a nonspherical Gaussian basis set.
In order to compare the efficiency of our basis set with the best data 
available in the literature \cite{wolniewicz:1995}, we performed a calculation 
at the equilibrium distance for the ground state of $H_2$. 
In this calculation we include $42$ atomic basis functions constituting the $H_2^+$ orbitals with 
a magnetic quantum number equal to zero, $21$ with gerade and $21$ with ungerade 
parity, respectively. Hereby $24$ atomic basis functions were optimized in order to describe the 
ground state of the hydrogen atom. In order to describe the in-out correlation correctly, 
$18$ further atomic basis functions were optimized for excited states of the hydrogen atom. 
The in-out correlation takes into account the fact that if one electron is 
located near the internuclear axis, the second electron is preferably 
located far from the internuclear axis. 
The angular correlation describing the fact that the angle between the two 
electrons preferably differs by $180$ degrees has been taken into account by using 
the following atomic basis functions: 
$32$ functions for each of the magnetic quantum numbers $m = +1$ and $m = -1$;
$18$ functions for $m = +2$ and $m = -2$; and analogously 
$12$ functions with $m = +3$ and $m = -3$. 
In order to understand the origin of the small difference of our dissociation 
energy compared to the dissociation energy given in Ref.\cite{wolniewicz:1995} 
we performed an additional calculation and included only the 42 functions described above with 
magnetic quantum number equal to zero. This yields the so-called $\sigma$ limit 
\cite{davidson:1962} of $1.1615\;a.u.$ which is the lowest 
energy achievable by molecular configurations including only functions with 
a magnetic quantum number equal to zero. 
Our calculation yields $1.16144\;a.u.$ for the $\sigma$ limit. 
Therefore the main difference between the benchmark energy for the ground state of 
the $H_2$ molecule and our energy is due to some missing in-out correlation and angular correlation. 

For the calculation of the entire PEC we used a second slightly smaller basis set 
than the one given above. The evaluation of the 
matrix elements takes increasingly more CPU time if higher angular momenta are involved. 
For the calculation of the PEC we therefore reduced the number of basis functions 
involving higher angular momenta in the following way: nine functions for 
each of the magnetic quantum numbers $m = +1$ and $m = -1$, and six functions for 
$m = +2$ and $m = -2$ were used. This reduction of our basis leads to only minor changes  
in the total energy but saves about $30$ \% of the CPU time needed for the 
evaluation of the matrix elements. 
In Table \ref{table1} both results for the ground state energy of the \wa are included. 
The calculations with respect to other electronic states of the \wa in field-free 
space were performed by using the second basis set described above. 
Details of the basis set can be obtained from the authors upon request. 

In contrast to the numerous investigations concerning the behavior and structure of the \wa 
in field-free space only a few studies deal with the \wa in strong magnetic fields
\cite{turbiner:1983,basile:1987,monteiro:1990,korolev:1992,lai:1992,demeur:1994,ortiz:1995,lai:1996}. 
Most of these deal with the \wa in superstrong magnetic fields as large 
$10^{11}$ or even $10^{12}\;G$ (in atomic units this corresponds to
$B  = 42.54414\;a.u.$ and $B  = 425.4414\;a.u.$, respectively). 

First of all we mention that our computational method is by no means restricted 
to a special range of the magnetic-field strength. We were therefore  
able to study the development of the total energy with respect to the field 
strength ranging from field-free space up to a very strong field. 
Before entering into a discussion of our results let us introduce our 
notation for the united and separated atom limit in the presence 
of a magnetic field. Throughout this paper we will denote the atomic hydrogen states in the 
dissociation limit with $H\left(m_a^{\pi_a}\right)$, where $m_a $ 
denotes the atomic magnetic quantum number and $\pi_a$ the atomic $z$ parity, 
respectively. The united atom limit is described by 
$^{2S+1} L_z^{\pi_z} $. Here $2S+1 $ is the spin multiplicity, $L_z $ is the projection of 
the electronic angular momentum onto the axis of the magnetic field and $\pi_z $ is the $z$ parity. 
Now we are in the position to discuss the structure of the electronic PEC 
for the $^1\Sigma_g$ state in the presence of a magnetic field. 

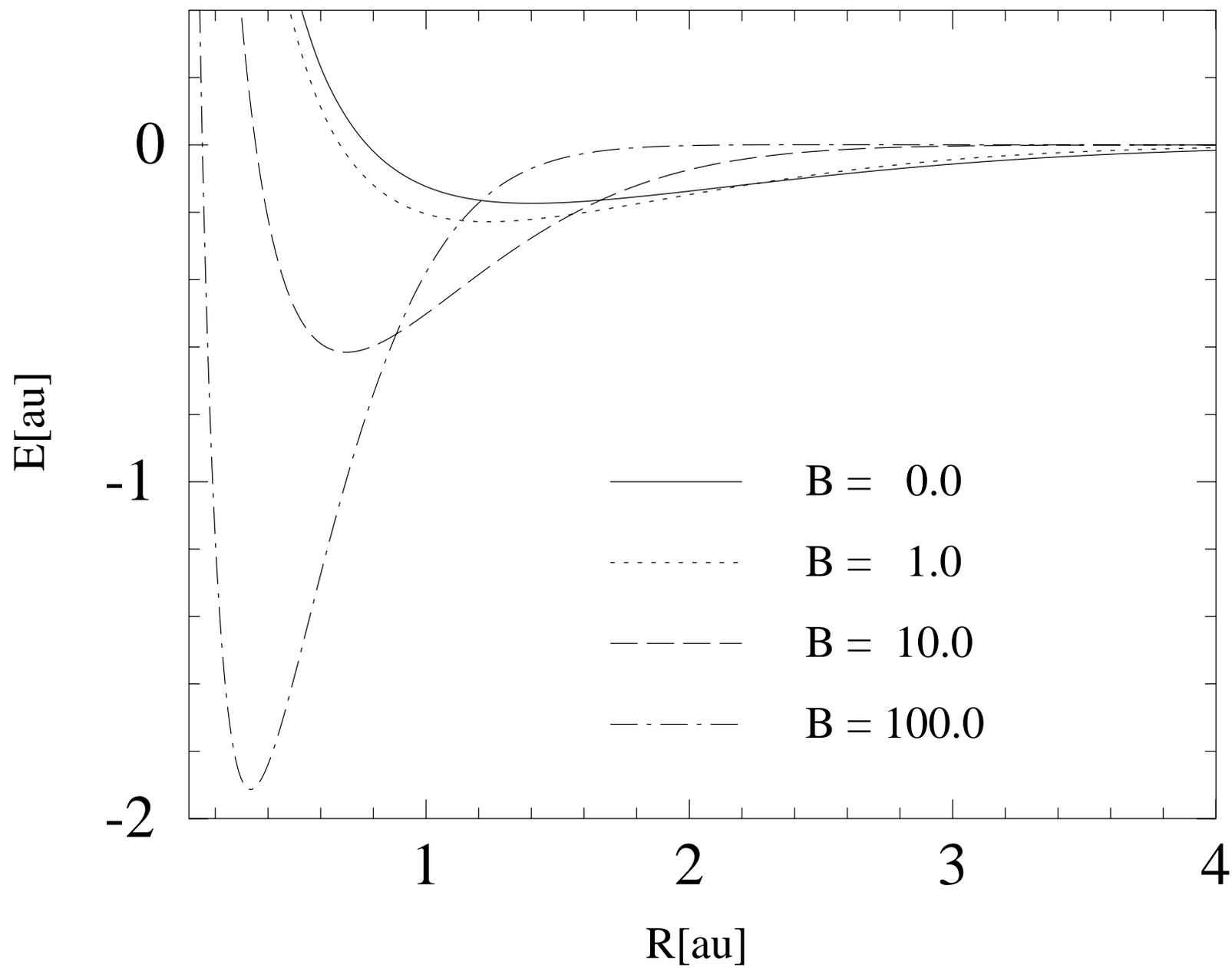
\begin{figure}
\caption{PECs for $B = 0.0\,,\,1.0\,,\,10.0$ and $100.0\;a.u.$ 
for the lowest $^1\Sigma_g$ state ; the energy is 
shown with respect to the dissociation limit i.e., 
$E\left(R\right) = E_t\left(R\right) - \lim\limits_{R\to\infty} E_t\left(R\right)$}
\label{fig1}
\end{figure}

Figure \ref{fig1} shows the energy curves of the $^1\Sigma_g$ state of the \wa for different 
field strengths. 
In order to display electronic energies for varying magnetic-field strengths in the same viewgraph 
the energy in the dissociation limit is substracted from the total energy, i.e., 
we show the quantity $E\left(R\right) = E_{t}\left(R\right) - \lim\limits_{R\to\infty} E_t\left(R\right)$. 
The total as well as the dissociation energies at the equilibrium distance of the $^1\Sigma_g$ state
are given in Table \ref{table1} together with the corresponding equilibrium internuclear 
distance for field strengths in the range $B = 0 - 100\;a.u.$. 
Furthermore, we show the total energy in the dissociation limit, i.e., 
$\lim\limits_{R\to\infty} E_t\left(R\right)$. 
For the $^1\Sigma_g$ state the dissociation channel is 
$H_2 \rightarrow H\left(0^+\right) + H\left(0^+\right)$ which means that the energy 
in the dissociation limit corresponds to the energy of two hydrogen atoms in the 
lowest electronic state with positive $z$ parity. 
The appropriate electronic state in the united atom limit is the $^10^+ $ 
helium state for any field strengths up to $B = 100 a.u.$. 

In the following we discuss the changes in the dissociation energy and equilibrium distance 
with increasing strength of the magnetic field. 
The overall behavior we observe in Table \ref{table1} is a monotonously increasing total energy 
as well as dissociation energy and a monotonously decreasing equilibrium internuclear distance. 
The decrease in the equilibrium internuclear distance 
originates from the simultaneous decrease of the electron clouds 
perpendicular and parallel to the magnetic field. 
Figure \ref{fig1} illustrates the drastic growth in the dissociation 
energy for magnetic-field strengths $B \lesssim 1\; a.u.$. At the same time the potential well  
becomes more and more pronounced, i.e., its width decreases strongly. 
Furthermore, the asymptotic behavior of the PEC for large values of $R$ changes 
with the magnetic-field strength. With an increasing value of $B$ the dissociation 
limit is reached at much smaller values of the internuclear distance, i.e 
the onset of the asymptotic behavior can be observed for much smaller internuclear distances. 
Furthermore transition states appear in the PEC for magnetic-field strengths $B\gtrsim1\;a.u.$. 
The corresponding data, i.e., the positions and total energies of the transition states with respect to 
the dissociation limit, can be found in Table \ref{table1}. 
For the position of the maximum we observe a strong decrease with increasing field strength. 
Simultaneously the height of the maximum with respect to the dissociation energy , i.e., 
$E_{t} - \lim\limits_{R\to\infty} E_{t}$, increases from $2\times10^{-6}$ at $B = 1.0\;a.u.$ to 
$6\times10^{-4}$ at $B = 100.0\;a.u.$. 

We emphasize that the $^1\Sigma_g$ state is not the 
ground state of the \wa in a magnetic field of arbitrary strength. 
In superstrong magnetic fields with $B \gtrsim 10^3 a.u. $ it is well known 
that the $^3\Pi_u$ state represents the ground state of the \wa \cite{ortiz:1995,lai:1995}. 
The ground state for magnetic-field strengths in the intermediate 
regime has not been investigated up to now, to our knowledge. For sufficienly weak 
magnetic fields the ground state has to be the $^1\Sigma_g$ state. 
However, we will show in section III.D that for magnetic-field strengths 
larger than $0.2\; a.u.$ or equal the lowest total energy of the $^3\Sigma_u$ state 
is lower than the lowest total energy of the $^1\Sigma_g$ state. This means that for 
$B \gtrsim 0.2\;a.u.$ the $^3\Sigma_u$ state is the ground state of the \wa. 
The ``crossing field strength'' 
of this lowest $^3\Sigma_u$ state with the $^3\Pi_u$ state, which is the ground state in 
superstrong magnetic fields, is not known so far. 

Now let us compare our results with the few data existing in the literature. 
Only two investigations were performed concerning the electronic structure of the \wa 
for magnetic-field strengths smaller than $B = 100\;a.u.$. 
In Ref. \cite{basile:1987}, PECs were shown for the $^1\Sigma_g$ 
state for magnetic-field strengths of $0.5\;a.u.$ and $1.0\;a.u.$. The authors used a very simple LCAO 
ansatz for the electronic wave function, including only one atomic orbital for the construction 
of the $H_2$ molecular wave function. The atomic orbitals had the following appearance: 
$\psi_\alpha\left(j\right) = \left(\frac{\epsilon ^2}{\pi}\right)^{1/2} 
exp\left[-\epsilon r_{\alpha j} \right]$ 
where $r_{\alpha j}$ is the distance of the $\alpha$ nucleus to the jth electron and 
$\epsilon$ is a variational parameter to be optimized. This ansatz is restricted to small 
magnitudes of the magnetic field ($B < 1\;a.u.$) where the spherical symmetry of the $^1\Sigma_g$ 
wave function is approximately conserved. 
For magnetic-field strengths of $0.5\;$ and $1.0\;a.u.$, the authors of Ref. \cite{basile:1987}  
provided an equilibrium distance of $1.29\;a.u.$
corresponding to a total energy of $-1.06\; a.u.$ and an equilibrium distance of $1.17\; a.u.$
with a total energy of $-0.83\; a.u.$, respectively. 
As shown in Table \ref{table2} we were able to improve these results significantly. 
The improvement in the total energy compared to the calculations of 
Ref. \cite{basile:1987} is approximately twice as large 
for $B = 1.0\; a.u.$ as for $B = 0.5\; a.u.$. 
For $B = 1.0\; a.u.$ the wave function of the $^1\Sigma_g$ state is deformed from a 
spherical one to a cylindrosymmetric one, and, therefore, a use of only one atomic orbital with 
a spherical symmetry is insufficient even for a qualitative description. 
In the second investigation performed by Turbiner \cite{turbiner:1983}, one of the simplest 
wave functions in a zero-order approximation was taken for the description of the 
$^1\Sigma_g$ state (see also Ref. \cite{dyson:1952}). 
In Table \ref{table2} we also compare our data with that of Ref. \cite{turbiner:1983}. 
We observe an overall improvement of the total energy in our results 
compared to the data obtained by Turbiner. 
For a magnetic-field strength of $5.0\times10^9\;G$ the equilibrium distance 
obtained by Turbiner differs significantly from our value leading to a relative 
difference in the total energy of $41$\%. 
We remark that the accuracy of the calculations in Ref. \cite{turbiner:1983} 
increases with increasing magnetic-field strength. For $B = 1.0\times10^{11}\;G$
the relative improvement in the total energy obtained by our calculations  
amounts to only about $0.5$ \% compared to \cite{turbiner:1983}. 

Finally let us investigate the question of whether vibrational levels exist 
in the PECs discussed above. This question is of 
great importance to the existence of bound states. 
In the presence of a magnetic field the determination of 
vibrational levels is a much more complicated task than in field-free space. 
First we note that the \bo known from field-free space is not valid in the 
presence of a magnetic field. In the presence of a magnetic field the nuclear charges 
are partially screened by the electrons against this field. In order to describe the partial 
screening of the nuclear charges against the magnetic field correctly, 
the diagonal term of the nonadiabatic 
coupling elements has to be included in the nuclear equation of motion. This leads to a new 
kind of adiabatic approximation, the partially screened \bo 
\cite{detmer:1995,schmelcher:1994,schmelcher2:1988}. 
Furthermore, the nuclear equation of motion explicitly not only depends on the internuclear distance 
but also on the angle between the internuclear axis and the magnetic-field axis. 
The facts discussed above clearly show that the nuclear dynamics is in general very complex. 
Within the present framework of the parallel configuration we can therefore provide only 
estimations of the energy levels of the vibrational states. Nevertheless this 
allows us to decide whether or not we encounter physically bound states with 
respect to the vibrational mode $R$. 
A lower bound of the vibrational energy within our approximation is given by the 
lowest vibrational state in the corresponding PEC using the field-free kinetic energy 
$\frac{\boldmath{P}^2}{2\mu}$. 
The corresponding nuclear equation of motion in field-free space for the given 
electronic PEC was solved with the help of a discrete variable representation \cite{colbert:1992}. 
The upper bound for the energy of the lowest vibrational state in the presence of a magnetic field 
was obtained by simply adding the Landau energy of the nuclear motion to the value of the 
energy obtained for the corresponding vibrational level in field-free space. 
In this way we obtained both upper and lower bounds for the vibrational levels. 
These estimations of the vibrational levels were performed for each PEC shown in 
Figure \ref{fig1} as well as the following figures. 

For the PEC of the $^1\Sigma_g$ state the procedure described above yields many, 
i.e., of the order of magnitude of a few dozens, of vibrational levels for the entire regime 
$B = 0-100\;a.u.$ of field strengths. This means that the $^1\Sigma_g$ state is 
a bound state with respect to the internuclear distance $R$ for this wide range of field strengths. 

\subsection{The $^1\Sigma_u$ state} 
The electronic PEC for the $^1\Sigma_u$ state in field-free space was 
calculated with high accuracy by Kolos and Wolniewicz \cite{kolos:1966,kolos:1968}. 
The energy curve in field-free space for this state possesses a minimum at an internuclear distance 
of $R = 2.43\; a.u. $. 
A detailed analysis of the wave function \cite{kolos:1966} shows the predominantly 
ionic character of the wave function for $3 < R < 7\;a.u.$. This ionic character has also been 
confirmed by an analysis of the corresponding rotation-vibration spectrum of the \wa 
\cite{herzberg:1959}. Since $H^-$ possesses only one weakly bound state in field-free 
space, one expects the ionic character of the hydrogen molecule in the $^1\Sigma_u$ state 
to decrease  with increasing internuclear distance. 
For large internuclear distances the wave function 
can be described as a mixture of $\left(1s\sigma2s\sigma\right)$ and
$\left(1s\sigma2p\sigma\right)$ configurations, where the 
$\left(1s\sigma2p\sigma\right)$ configuration predominates for very large internuclear 
distances. Therefore, the dissociation channel is given by 
$H_2 \rightarrow H\left(1s\right) + H\left(2p\right)$. 
The corresponding state in the united atom limit is the electronic $^3P \: 1s2p $ helium state. 
In order to check the convergence of our calculations we compared our results with 
the data given in Ref. \cite{kolos:1968}. The overall relative accuracy in the total energy of the present 
calculation for the $^1\Sigma_u$ state in field-free space 
is better than $10^{-3}$. As can be seen from Table \ref{table3} our 
equilibrium distance differs by $0.01\;a.u.$ from the correct value of $R = 2.43\;a.u.$. 

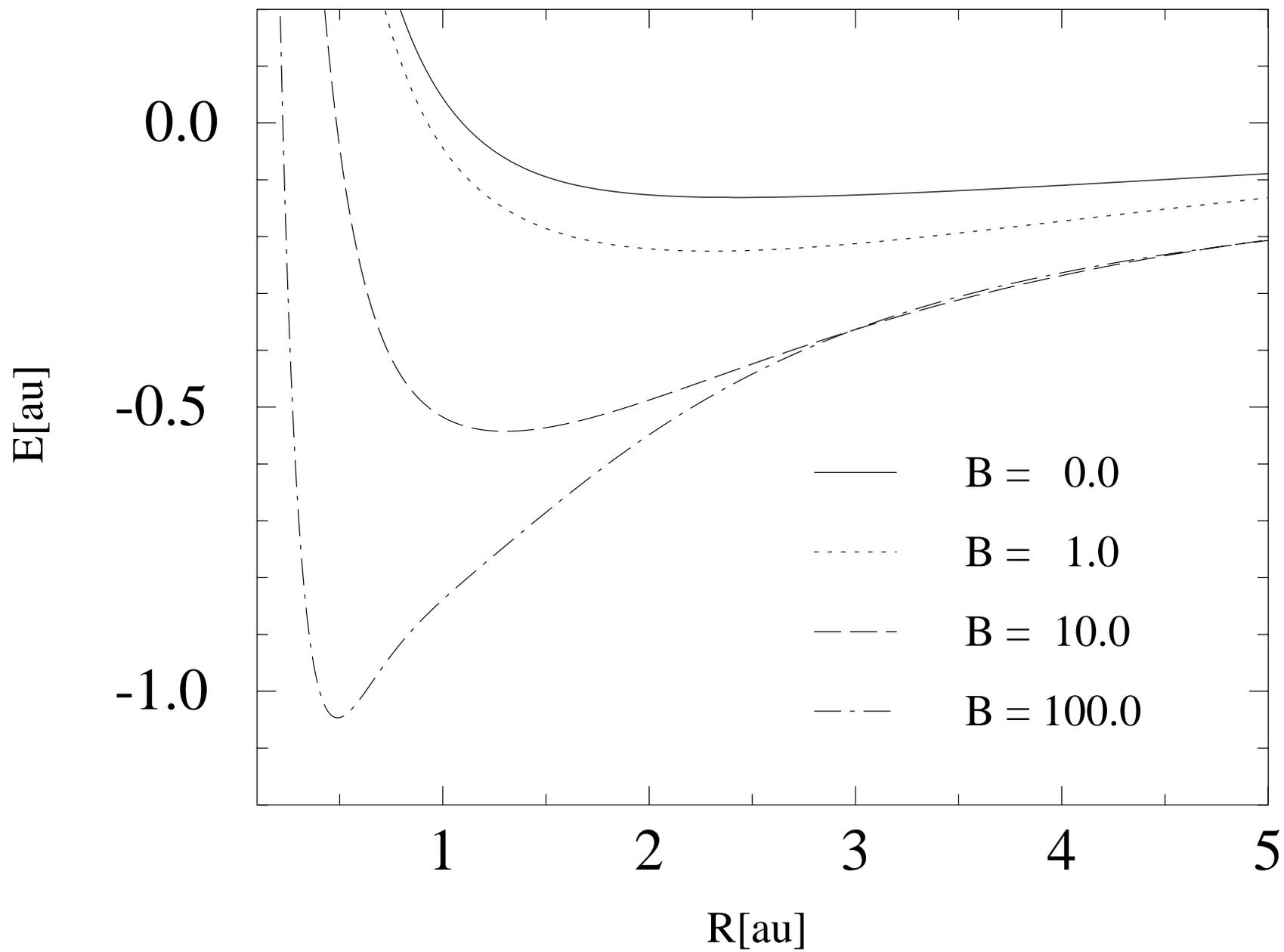
\begin{figure}
\caption{PECs for $B = 0.0\,,\,1.0\,,\,10.0$ and $100.0\;a.u.$ 
for the lowest $^1\Sigma_u$ state; the energy is 
shown with respect to the dissociation limit i.e., 
$E\left(R\right) = E_t\left(R\right) - \lim\limits_{R\to\infty} E_t\left(R\right)$}
\label{fig2}
\end{figure}

In the presence of a magnetic field we observe a monotonous increase in the total energy with 
increasing magnetic-field strength. The corresponding values are given in 
Table \ref{table3}. At the same time the dissociation energy increases 
monotonously (cf. Fig. \ref{fig2} and Table \ref{table3}). 
It can be seen from Table \ref{table3} that 
the value of the equilibrium distance exhibits a minor increase from $2.42\;a.u.$ in 
field-free space to $2.53\;a.u.$ for $B = 0.2\;a.u.$. However, 
for $B \ge 0.2\;a.u.$  we observe a drastic 
decrease in the equilibrium internuclear distance with increasing magnetic-field strength. 
 For $B = 1.0$ and $100\;a.u.$, the corresponding values are 
$R_{eq} = 2.30$ and $0.490\;a.u.$, respectively. 
Figure \ref{fig2} shows the well becoming more and more pronounced 
with increasing magnetic-field strength while the onset of the asymptotic convergence 
behavior is shifted to larger internuclear distances with 
increasing field strength. 

Another important property of the $^1\Sigma_u$ state in a magnetic field 
is the change of the dissociation channel in sufficiently strong fields. 
As described above, the wave function in field-free space 
exhibits a partially ionic character for certain internuclear distances. 
However, the dissociation into $H^- + H^+$ is not possible due to the 
nonexistence of strongly bound electronic states of $H^-$. 
In a magnetic field, however, it is known that each negative ion possesses infinitely many bound states 
in a nonzero constant magnetic field \cite{avron:1981,lieb:1986}. This means we encounter 
an infinite number of bound states of $H^-$ in the presence of a magnetic field even though 
$H^-$ has only one bound state in field-free space. 
With increasing field strength the ground state energy of $H^-$ shows a 
monotonous behavior, and $H^-$ becomes more and more strongly bound. 
Consequently, there exists a critical field strength for which the 
total energy of the ground state of $H^-$ and the 
total energy of $H\left(0^+\right) + H\left(0^-\right)$, which is the dissociation limit 
for weaker field strengths, are equal. 

The above considerations help us to understand the asymptotic behavior in the dissociation 
limit of the $^1\Sigma_u$ state as well as the shape of the PEC: 
For magnetic-field strengths $B\lesssim10.0\;a.u.$ the molecular $^1\Sigma_u$ state 
dissociates into $H\left(0^+\right) + H\left(0^-\right)$. 
Since the total energy of $H^-$ is larger than the energy of $H\left(0^+\right) + H\left(0^-\right)$ 
and since no crossings are allowed between electronic states of the same symmetry 
an avoided crossing between the ground state and the first excited state in the $^1\Sigma_u$ 
subspace occurs in this regime of field strengths. 
With increasing magnetic-field strength $B=0\rightarrow10\;a.u.$ the position of the avoided crossing is 
shifted to increasingly larger internuclear distances (for $B = 0.1\;$ and $10.0\;a.u.$ 
the position of the crossing is at $\sim12$ and $81.47\;a.u.$, respectively). 
Between $B = 10.0$ and $50.0\;a.u.$ the dissociation channel 
changes from $H\left(0^+\right) + H\left(0^-\right)$ to 
$H_2 \rightarrow H^-\left(0_s^+\right) + H^+$, where the subscribt s denotes 
that the $H^-$ state is a singlet state. The ``transition field strength'' is approximately
$B\sim20\;a.u.$. Due to the change in the dissociation channel 
the onset of the asymptotic behavior of the PEC is shifted to increasingly larger internuclear 
distances. This leads to a totally different shape of the PEC which can be seen for 
$B = 100.0\;a.u.$ from Fig. \ref{fig2}. 
As a further result we obtain the ionization energy for $H^-$ 
for magnetic-field strengths of $B = 50.0$ and $100.0\;a.u.$. 
The resulting total binding energies with respect to both electrons are 
$3.637999\;a.u.$ and $4.561968\;a.u.$, respectively. The best available data for the 
ionization energy of the $H^--$ion in a strong magnetic field $B \ge 100\;a.u.$ are given in 
Refs. \cite{vincke:1989,larsen:1979}. Compared to the value given in Ref.  
\cite{vincke:1989} for $B = 100.0\;a.u.$ our result of $4.561968\;a.u.$ is more than $1$\% 
lower in energy. The ionization energy for $H^-$ at $B = 50.0\;a.u.$ amounts to $3.637999\;a.u.$ 
and no corresponding values have been known from the literature so far. 

Finally, we briefly comment on the existence of vibrational levels. 
Many vibrational levels were found to exist in the above PECs for arbitrary 
field strengths up to $100.0\;a.u.$. 
With increasing field strength the overall tendency is an increase in the number of 
vibrational levels.

\subsection{The $^3\Sigma_g$ state}
The Born-Oppenheimer PEC for the lowest $^3\Sigma_g$ state has been calculated 
with very high accuracy in field-free space. Accurate studies have been performed 
using explicitly correlated methods, which include wave functions of generalized 
James-Coolidge or Kolos-Wolniewicz type \cite{kolos1:1975,kolos:1978,bishop:1981} and  
full CI calculations with large elliptical basis sets \cite{liu:1994}. 
Our CI calculation yields $0.737124\;a.u.$ for the total energy at the equilibrium distance 
of $R = 1.87\;a.u. $ (see Table \ref{table4}). 
The relative accuracy compared to the best data available 
in the literature \cite{bishop:1981} amounts to $4.4\times10^{-5}$. 
We emphasize that this high precision 
of our data is obtained for arbitrary internuclear distances. 
The dissociation channel for the $^3\Sigma_g$ state in field-free space is 
$H_2 \rightarrow H\left(1s\right) + H\left(2s\right)$ and the helium state in the 
united atom limit is the $^3S\;1s2s$ state. 

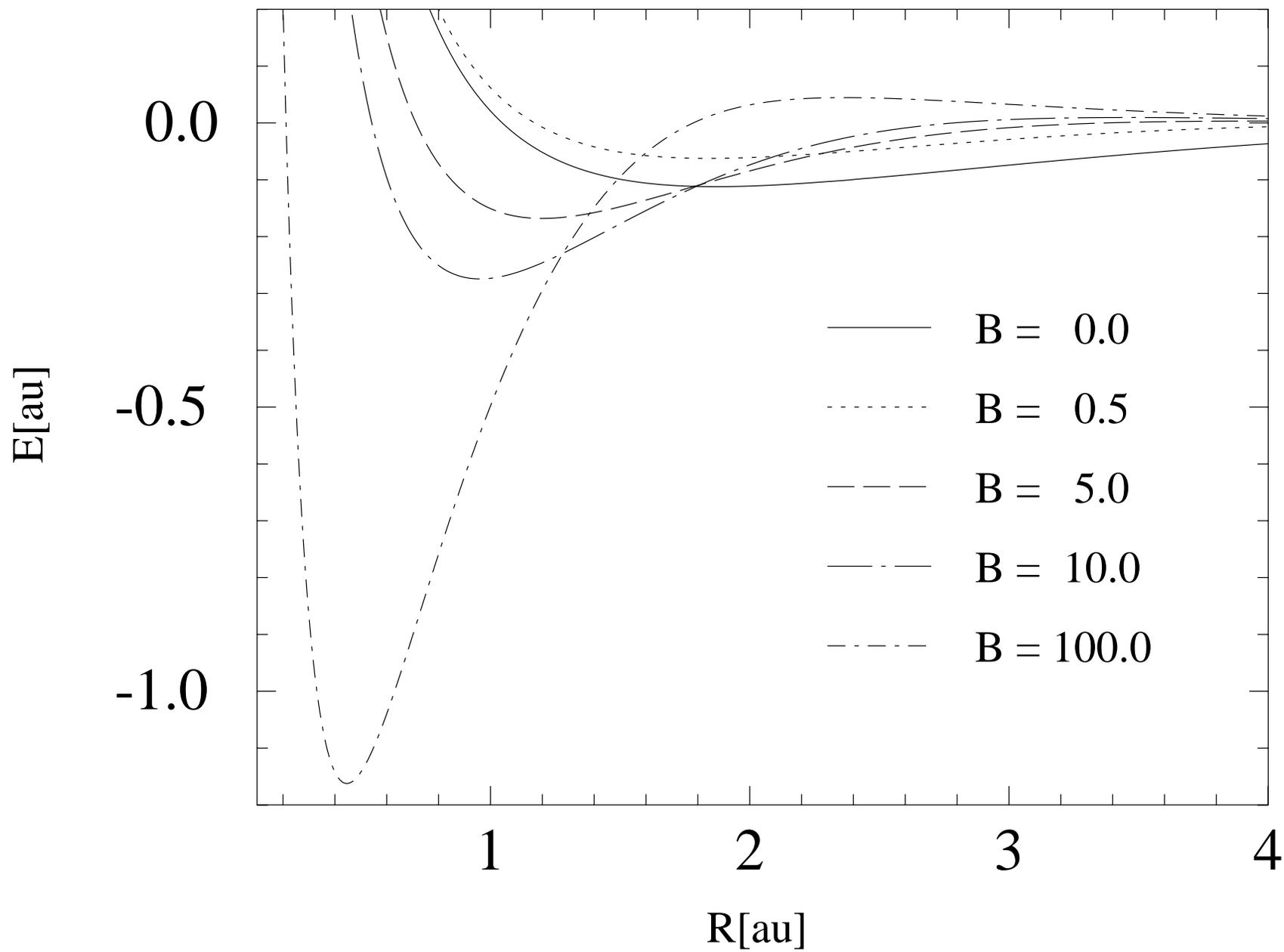
\begin{figure}
\caption{PECs for $B = 0.0\,,\,0.5\,,\,5.0\,,\,10.0$ and $100.0\;a.u.$ in the range $0.1 < R < 4\;a.u.$
for the lowest $^3\Sigma_g$ state showing the first minimum of this state 
; the energy is 
shown with respect to the dissociation limit i.e., 
$E\left(R\right) = E_t\left(R\right) - \lim\limits_{R\to\infty} E_t\left(R\right)$}
\label{fig3}
\end{figure}

In the presence of a magnetic field the $^3\Sigma_g$ state dissociates into 
$H\left(0^+\right) + H\left(0^-\right)$. The corresponding united atom state is the $^30^+ $ 
helium state. Due to the spin-Zeeman shift the total energy monotonously decreases with 
increasing magnetic-field strength as can be seen in Table \ref{table4}. 
For magnetic-field strengths below $0.5\;a.u. $ we observe a 
monotonous decrease in the dissociation energy $E_{d1}$ with 
increasing magnetic-field strength 
from $0.112124$ to $0.062317\;a.u.$ for $B = 0.0$ and $0.5\;a.u.$, respectively. 
At the same time the equilibrium distance decreases slightly from $1.87\;a.u.$ to $1.81\;a.u.$. 
For magnetic-field strengths larger than $0.5\;a.u. $ we observe a drastic 
increase in the dissociation energy and a simultaneous decrease in the equilibrium internuclear distance. 
Detailed numerical data for the PEC are provided in Table \ref{table4} and 
the development of the electronic PEC for internuclear 
distances smaller than $4.0\;a.u.$ is illustrated in Fig.\ref{fig3}. 
In Fig. \ref{fig3} we can observe that the well of the PEC becomes increasingly more shallow 
for increasing field strengths in the range $B = 0.0 \rightarrow 0.5\;a.u.$. 
With further increasing magnetic-field strength the opposite behavior can be observed and 
the potential well becomes more and more pronounced and deeper. 

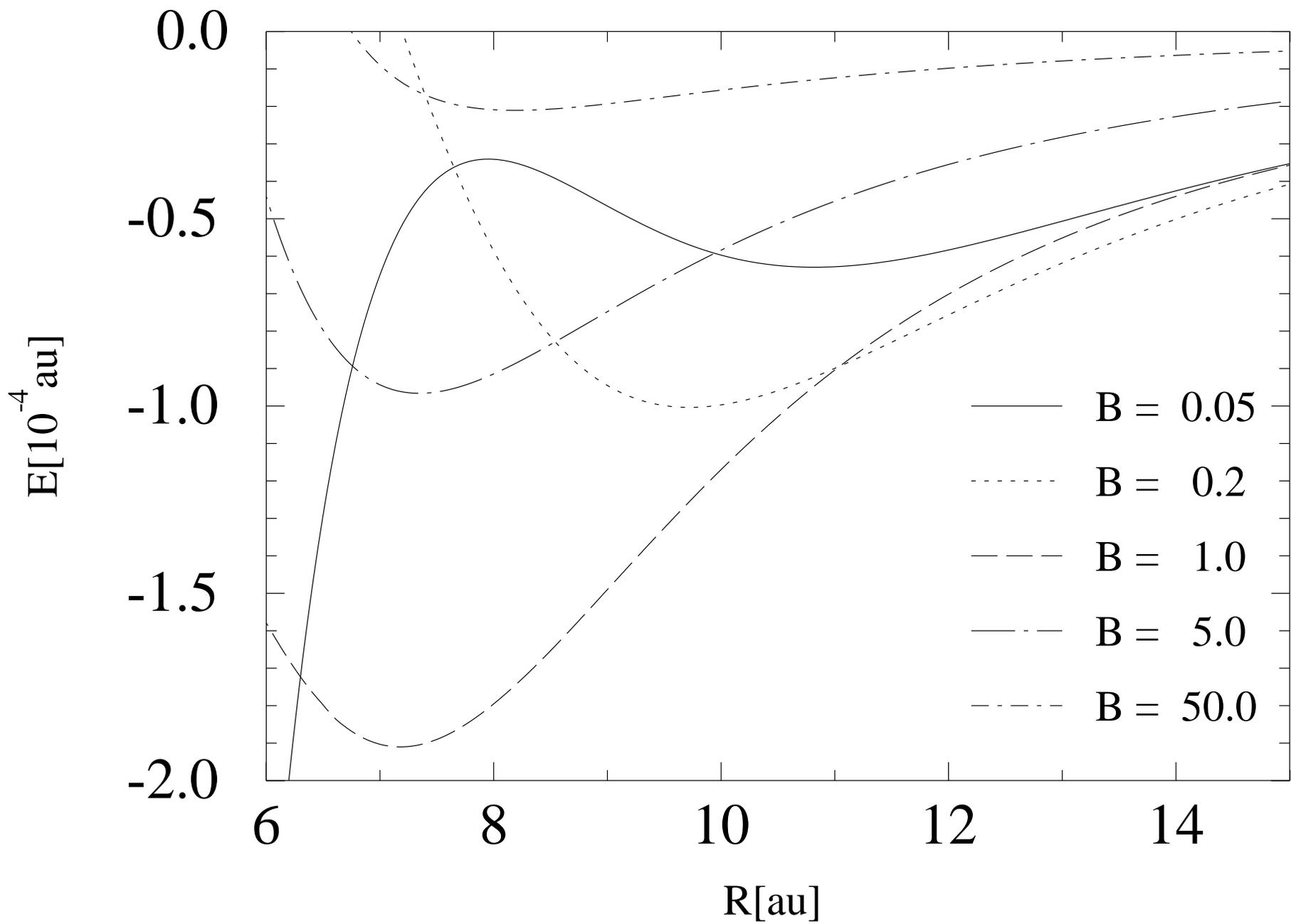
\begin{figure}
\caption{PECs for $B = 0.05\,,\,0.2\,,\,1.0\,,\,5.0$ and $50.0\;a.u.$ in the range $6 < R < 15\;a.u.$ 
illustrating the second minimum of the lowest $^3\Sigma_g$ state; the energy is 
shown with respect to the dissociation limit i.e., 
$E\left(R\right) = E_t\left(R\right) - \lim\limits_{R\to\infty} E_t\left(R\right)$}
\label{fig4}
\end{figure}

A closer look at the PEC 
reveals that for magnetic-field strengths between $0.05$ and $50.0\;a.u.$ the $^3\Sigma_g$ state 
develops a second minimum, which has no counterpart in field-free space. However, this minimum 
is very shallow and the maximum dissociation energy amounts to only $1.91\times10^{-3}\;a.u.$ 
at a magnetic-field strength of $1.0\;a.u. $. From Table \ref{table4} 
we see the dependence of the dissociation energy and the equilibrium internuclear 
distance of this second minimum on the magnetic-field strength. 
The corresponding PECs are shown in Fig.\ref{fig4}. 
In this figure we observe a hump of the $^3\Sigma_g$ state at $B = 0.05\;a.u.$. This 
hump also occurs for higher magnetic-field strengths, but is shifted to smaller internuclear 
distances and, therefore cannot be seen in Fig. \ref{fig4}. 
For the location of the maximum we observe an overall decrease in the internuclear distance 
with increasing magnetic-field strength. Numerical data concerning the position and 
total energy of the maximum in the energy curve are also given in Table \ref{table4}. 
Both the maximum and second minimum of the PEC appear at the same magnetic-field strength of $0.05\;a.u.$.  
However, the second minimum vanishes for magnetic-field strengths larger than $50.0\;a.u.$ 
whereas the hump remains in the PEC. 

Let us finally focus on the existence of vibrational levels. 
For the first minimum we found in the order of ten vibrational levels. 
The number of levels slightly decreases from $B = 0.0$ to $0.5\;a.u.$ with decreasing 
depth of the well. For larger field strengths ($5.0,10.0$, and $100.0\;a.u.$) 
the number of levels slightly increases again. For the second, i.e., outer, 
minimum we found about five vibrational levels for the 
field strengths $0.05$, $0.2$, and $1.0\;a.u.$, respectively. 
For a field strength $B = 5.0\;a.u.$ the lower bound of the vibrational energy 
lies inside the potential well whereas the upper bound is above the well. 
Here the detailed dynamical behavior decides on the existence of vibrational levels, and 
no definite conclusion can be drawn from our estimations. 
For $B = 50.0\;a.u.$ the same conclusion holds. 

\subsection{The $^3\Sigma_u$ state}
In field-free space the electronic PEC of the $^3\Sigma_u^+$ state 
is repulsive, i.e., does not exhibit a well-pronounced potential well. 
The united atom limit of this state is the $^3S\;1s2p$ helium state, and the dissociation 
channel is $H_2 \rightarrow H\left( 1s \right) + H \left( 1s \right)$. 
One of the early but very accurate Born-Oppenheimer calculations on the lowest $^3\Sigma_u^+$ state 
was performed by Kolos and Wolniewicz \cite{kolos:1965,kolos:1974}. 
Recently several calculations were performed in order to improve the Born-Oppenheimer potential 
energy for this state \cite{liu:1994,liu:1993,frye:1989,borondo:1986}. 
Most accurate results were obtained using a Hylleraas-type expansion \cite{kolos:1965,kolos:1974},  
explicitly correlated Cartesian Gaussian basis functions \cite{frye:1989}, 
or elliptical basis functions \cite{liu:1994}. 

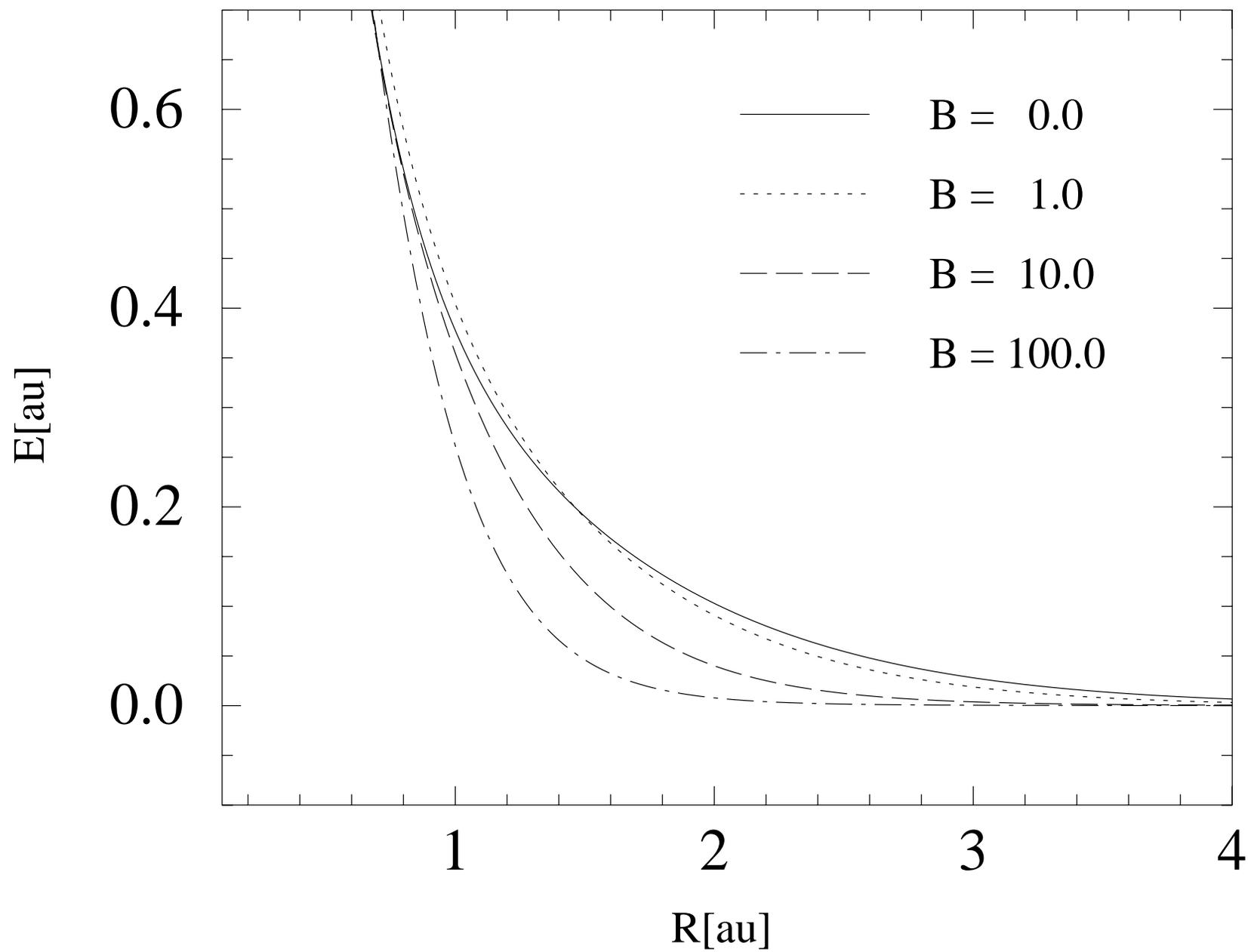
\begin{figure}
\caption{PECs for $B = 0.0\,,\,1.0\,,\,10.0$ and $100.0\;a.u.$ for the 
lowest $^3\Sigma_u$ state; the energy is 
shown with respect to the dissociation limit i.e., 
$E\left(R\right) = E_t\left(R\right) - \lim\limits_{R\to\infty} E_t\left(R\right)$}
\label{fig5}
\end{figure}

As a reference for our calculations we used the data given in Ref. \cite{liu:1994} 
for internuclear distances smaller than $4.0\;a.u.$ and 
Refs. \cite{kolos:1965,kolos:1974} for $R \ge 4.0\;a.u.$.
For $R \le 4 a.u.$ we obtained an overall relative accuracy of $6\times10^{-4}$ compared to the 
results in Ref. \cite{liu:1994}. For larger values of the internuclear distance this 
accuracy further increases, and we obtained a relative accuracy of at least $3\times10^{-6}$. 
The best conventional CI calculations for this state with a basis set of spherical 
Gaussian-type orbitals were 
performed in Ref. \cite{borondo:1986}. Comparing our results 
with the data given in Ref. \cite{borondo:1986} it can be seen that our values of the 
total energy are lower by $1.0$\% and $0.4$\% for $R \le 2.0 $ and 
$R \ge 2.0 $, respectively. Our accurate results, in particular for the $^3\Sigma_u^+$ state 
in field-free space demonstrate the usefulness of our basis set of nonspherical nonorthogonal 
Cartesian Gaussian basis functions. The PEC for the $^3\Sigma_u^+$ state 
for different magnetic-field strengths is presented in Fig.\ref{fig5}. 

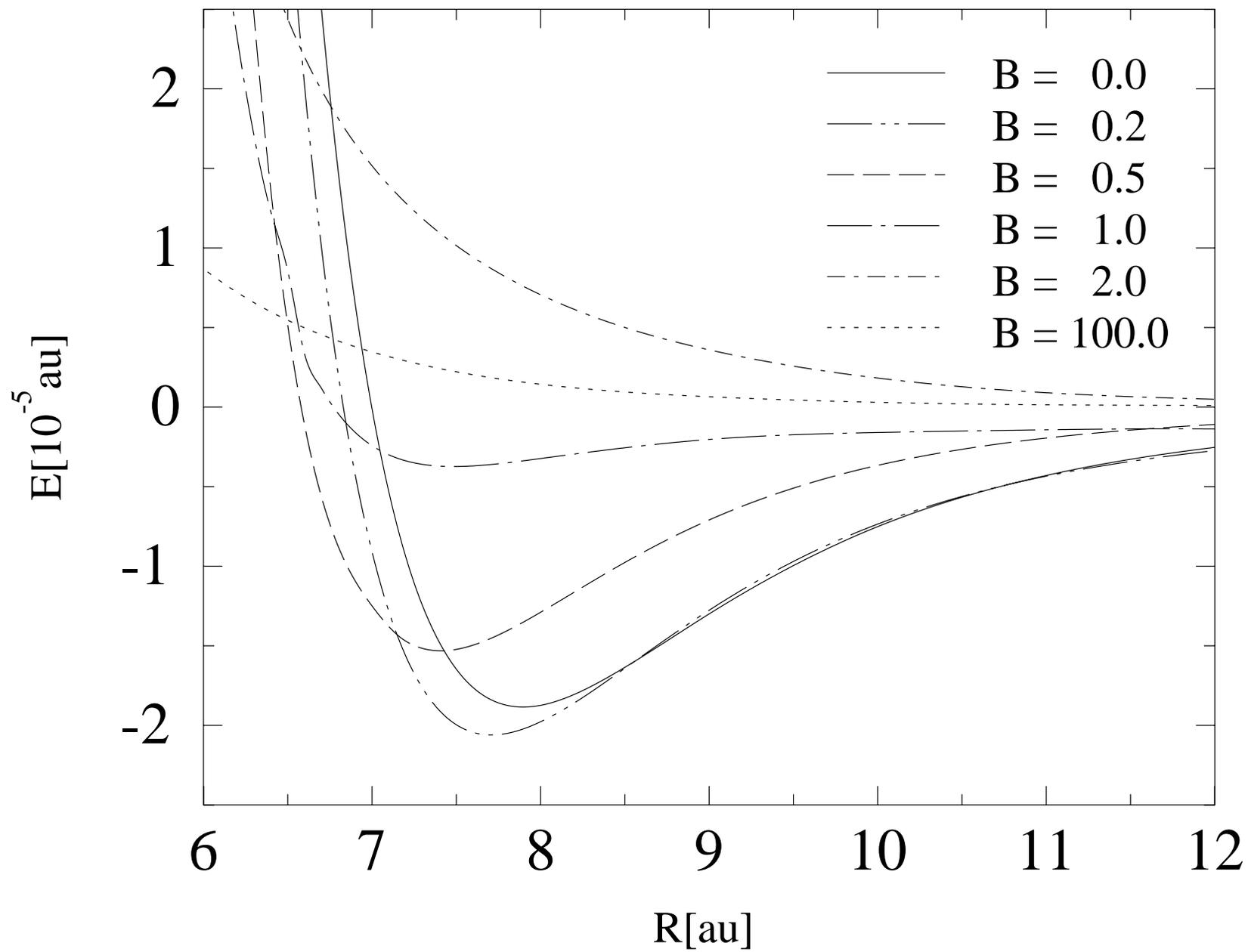
\begin{figure}
\caption{van der Waals minima of the lowest $^3\Sigma_u$ state for 
$B = 0.0\,,\,0.2\,,\,0.5\,,\,1.0\,,\,2.0$ and $100.0\;a.u.$; the energy is 
shown with respect to the dissociation limit i.e., 
$E\left(R\right) = E_t\left(R\right) - \lim\limits_{R\to\infty} E_t\left(R\right)$}
\label{fig6}
\end{figure}

Despite the fact that the PEC of the $^3\Sigma_u^+$ state 
is predominantly repulsive it exhibits a very shallow van der Waals minimum 
around $R \sim 8\;a.u. $. Due to the dissociation of the $^3\Sigma_u^+$ state into 
$H \left( 1s \right) + H \left( 1s \right)$, we encounter a dipole-dipole interaction 
of induced dipole moments for large internuclear distances which is proportional to $\frac{-1}{R^6}$.  
The PEC in the vicinity of the van der Waals minimum 
is presented in Fig.\ref{fig6}, and numerical data are given in Table \ref{table6}.  
For the van der Waals minimum we obtained a dissociation energy of $1.885 \times 10^{-5}\;a.u.$ 
at an internuclear distance of $7.9\;a.u.$. 
The determination of the van der Waals energy and the corresponding minimum 
is a delicate task and requires very large basis sets as well as high numerical accuracy. 
Within our calculations the position of the minimum has been determined with 'only' two 
significant digits. Since the dissociation energy is very small the relative accuracy of our result
of the dissociation energy differs from the high precision calculations in Refs.  
\cite{kolos:1965,kolos:1974} by about $17 \%$. We emphasize that the description of 
the van der Waals minimum is a drastic improvement compared to the 
results of Ref. \cite{borondo:1986} where no minimum was predicted. 

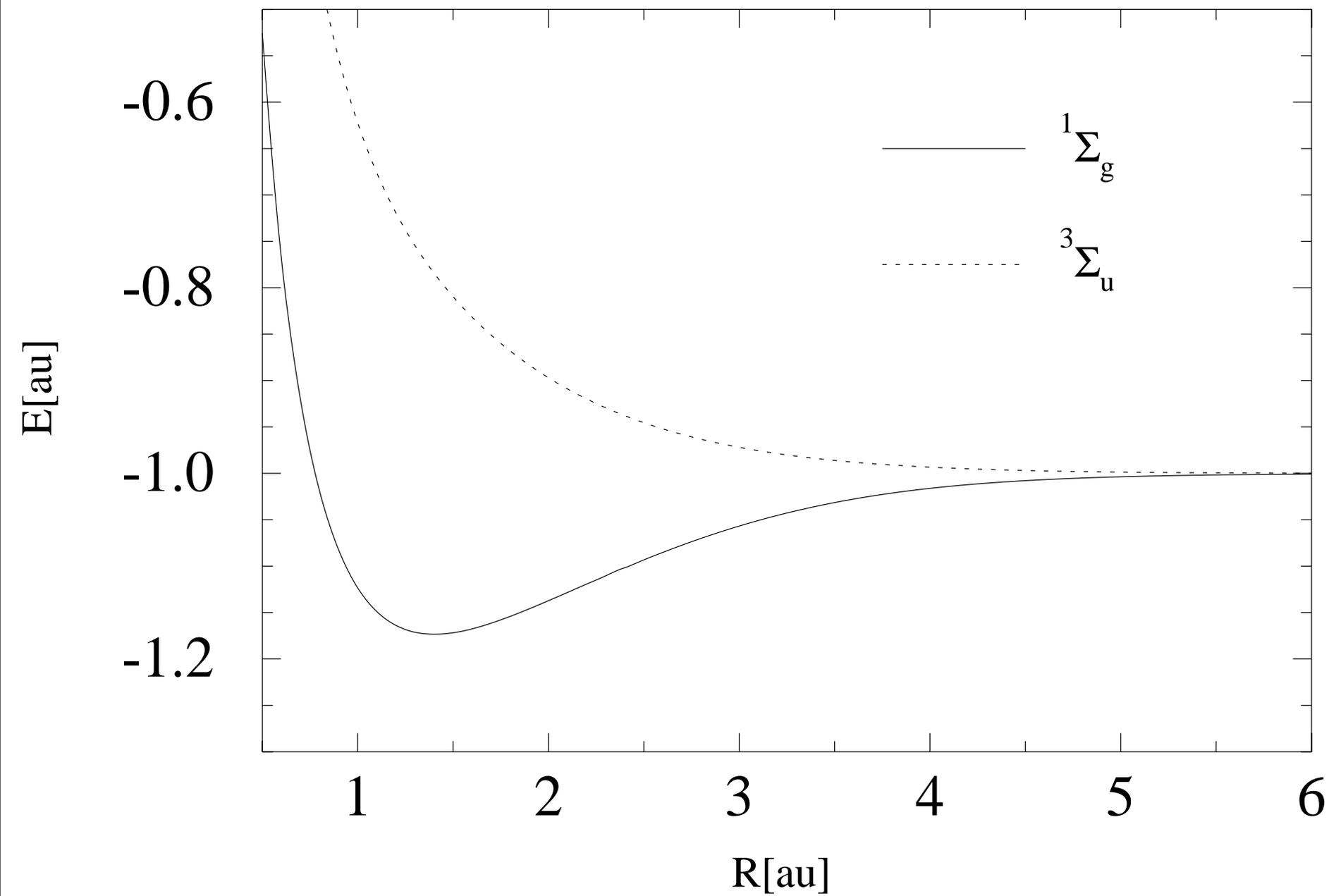
\begin{figure}
\caption{Total energy of the $^1\Sigma_g$ and $^3\Sigma_u$ state for $B = 0.0\;a.u.$ 
(all quantities are given in atomic units)}
\label{fig7}
\end{figure}

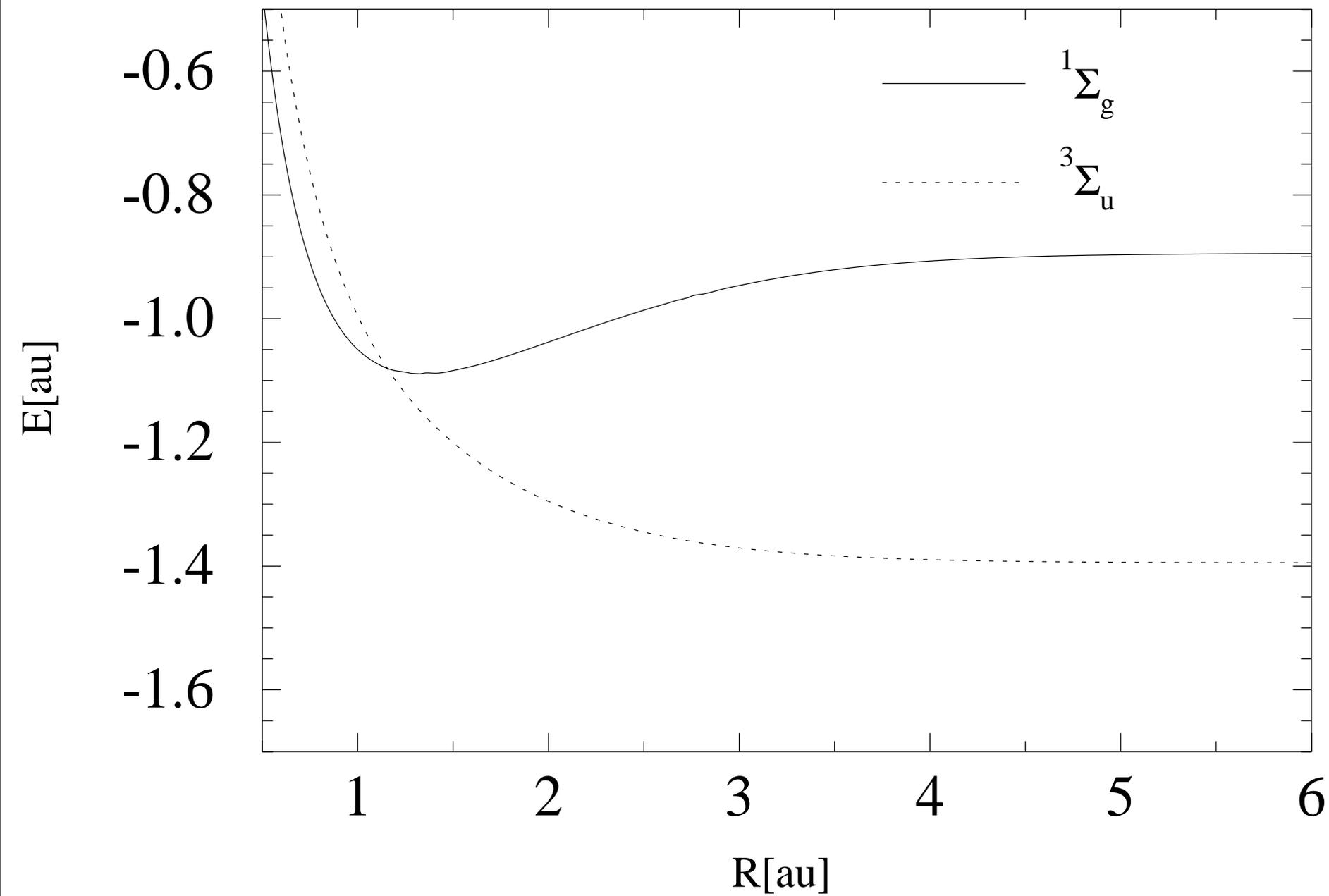
\begin{figure}
\caption{Total energy of the $^1\Sigma_g$ and $^3\Sigma_u$ state for $B = 0.5\;a.u.$ 
(all quantities are given in atomic units)}
\label{fig8}
\end{figure}

In the following we discuss the development of the PEC 
for the $^3\Sigma_u^+$ state depending on the magnetic-field strength. 
First, we focus on the global structure which is shown in Fig. \ref{fig5}. 
In the presence of a magnetic field the separated atom limit is given by 
$H_2 \rightarrow H\left(0^+\right) + H\left(0^+\right)$, i.e., the molecule dissociates 
into two hydrogen atoms in their ground states with positive $z$ parity. 
The corresponding united atom state in the presence of a magnetic field is the $^30^- $ helium state. 
First of all we mention that the onset of the asymptotic behavior with respect to the dissociation 
occurs for increasingly smaller internuclear distances with increasing magnetic-field strength. 
A comparison of the energy values given in Tables \ref{table1} and Table \ref{table6} 
reveals that for magnetic-field strengths larger than $0.2\;a.u.$ 
the $^3\Sigma_u$ state is lower in energy than the $^1\Sigma_g$ state. Therefore, the crossing 
between these two states happens between the two field strengths $0.1$ and $0.2\;a.u.$. 
In Fig. \ref{fig7} and Fig. \ref{fig8} we illustrate the crossing of these two states. 
In Fig. \ref{fig7}, which shows the total energy of the $^1\Sigma_g$ and $^3\Sigma_u$ states at 
a magnetic-field strength of $B = 0.0\;a.u.$, we can see the $^3\Sigma_u$ state being higher in 
the total energy as the $^1\Sigma_g$ state. Figure \ref{fig8} shows the same 
states at a magnetic-field strength of $B = 0.5\;a.u.$. In this figure we observe a drastic 
decrease in the total energy of the $^3\Sigma_u$ state. 
The crossing of these two states has an important concequence for the stability 
of molecular hydrogen in astrophysics. Beyond $B \sim 0.2\;a.u.$ the ground 
state of the hydrogen molecule is the $^3\Sigma_u$ state which is an unbound 
electronic state at least for the parallel configuration. 
It is a challenging task to clarify whether this is true for any angle of the 
internuclear axis with respect to the magnetic-field axis.  
In principle it is possible that a potential well might develop if the internuclear 
and magnetic-field axes do not coincide. The investigation of such configurations is an 
important task in the future. In addition we mention that beyond $B \gtrsim 3\times10^3\;a.u.$ 
the global ground state of $H_2$ is the $^3\Pi_u$ state \cite{lai:1992,ortiz:1995}. 

In the following we investigate the development of the van der Waals minimum 
depending on the magnetic-field strength which is of particular interest for the 
determination of the global ground state of the \wa in a magnetic field. 
In field-free space the van der Waals potential is given
by a law proportional $\frac{1}{R^6}$ due to the dipole-dipole interaction of 
induced dipoles in first order perturbation theory. 
In the presence of a magnetic field we have to pay attention to another interaction between 
atoms in s states. In first-order perturbation theory two atoms in a magnetic field 
interact like two permanent quadrupoles. Therefore the leading expression in first-order perturbation 
theory is proportional to $\frac{1}{R^5}$.  
In Fig. \ref{fig6} we show the development of the van der Waals minimum with increasing 
magnetic-field strength. As can be seen from Table \ref{table6} the dissociation energy 
increases by $0.175 \times 10^{-5} a.u.$ if we increase the field strength from zero to $0.2\;a.u.$.  
At the same time we observe a monotonous decrease of the internuclear distance 
corresponding to the minimum from $7.9$ to $7.7\;a.u.$. 
In Fig. \ref{fig6} we observe that the shape of the energy curve 
for $B = 0.2\;a.u.$ differs only slightly from that in field-free space. 
For magnetic-field strengths larger than $B =\;0.2$ the dissociation energy decreases drastically 
with increasing magnetic-field strength down to $0.375 \times 10^{-5}$ for $B = 1.0\;a.u.$. 
Simultaneously the energy curve changes its shape. 
The gradient of the energy with respect to internuclear distances $R \le R_{eq}$ 
becomes larger and a more and more shallow minimum can be observed. 
For magnetic-field strengths larger than $1.0\;a.u.$ no minimum is observed. 
The gradient of the energy with respect to the internuclear distance increases further 
with increasing magnetic-field strength, changing the appearance of the PEC 
to an increasingly flatter curve. 
No vibrational levels were found for any field strength. 

Finally we draw the readers attention to the fact that the 
calculations concerning the position and dissociation energy of the van der Waals minimum 
are close to the convergence limit of our calculations. The disappearance of the van der Waals 
minimum within our calculations to our opinion reflects a real physical effect, but 
needs further investigation for a definite clarification. 
In order to answer the question about the lowest bound state 
of the \wa in the high field regime, a detailed 
and very accurate investigation of the $^3\Pi_u$ electronic state has 
to be performed.

\section{Summary and Conclusions}
In the present paper we investigated the electronic structure of the \wa in a magnetic field. 
We hereby focused on the case of a parallel internuclear axis and a magnetic-field axis 
for a magnetic quantum number equal to zero. The key ingredient for our CI calculations is 
a basis set of nonorthogonal nonspherical Gaussian atomic orbitals which was 
established previously. The corresponding basis sets can be obtained from the authors 
upon request. Our results for the PECs in field-free space 
show a high accuracy compared to the existing data in the literature. 
The non-sherical atomic orbitals may therefore be very useful for precision calculations concerning 
molecules in field-free space. 

First we investigated the lowest $^1\Sigma_g$ state which is the ground state 
of the \wa in field-free space. In the presence of a magnetic field we observed 
a monotonous increase in the total energy. At the same time the equilibrium distance 
decreases and the dissociation energy (chemical binding energy) 
increases rapidly. The few existing data concerning 
the total energy of the $^1\Sigma_g$ state in the presence of a magnetic field were 
significantly improved by our calculations. 
By calculating lower and upper bounds for the lowest vibrational energy the PEC of the 
$^1\Sigma_g$ state was shown to contain many vibrational levels for any magnetic-field strength up to 
$100\;a.u.$. 

In the next step we studied the lowest state in the $^1\Sigma_u$ subspace. 
With increasing magnetic field we first observed a minor increase in the equilibrium 
internulear distance in the range $0.0 \le B \le 0.2\;a.u.$. 
The dissociation and the total energy increases  
monotonously with increasing magnetic-field strength. 
Many vibrational levels were found for the PEC of the $^1\Sigma_u$ state in the 
entire regime $B = 0-100\;a.u.$. The number of levels hereby increase 
with increasing field strength. 
An important feature of the $^1\Sigma_u$ 
state is the change in the dissociation channel with increasing magnetic-field strength. 
In field-free space we have $H_2 \rightarrow H\left(1s\right) + H\left(2p\right)$ in the 
separated atom limit. The wave function possesses a predominantly ionic character for 
large values of the internuclear distance. However, in field-free space the dissociation 
$H_2 \rightarrow H^- + H^+$ of the lowest $^1\Sigma_u$ state is not possible due 
to the nonexistence of strongly bound states 
for the $H^--$ion in field-free space. In contrast to this strongly bound states of $H^-$ exist 
in the presence of a magnetic field. Therefore a dissociation into $H^- + H^+$ is possible and we 
observe a change in the dissociation channel for magnetic-field strengths between 
$B = 10.0$ and $50.0\;a.u.$: For $B \le 10.0\;a.u.$ we have  
$H_2 \rightarrow H\left(0^+\right) + H\left(0^-\right)$. For field strengths slightly larger than 
$10.0\;a.u.$ the dissociation limit is given by $H_2 \rightarrow H^-\left(0_s^+\right) + H^+$. 
As a result of our calculations we therefore obtained the dissociation energy for 
$H^-$ for $B = 50.0$ and $100.0\;a.u.$. Our result of $4.561968\;a.u.$ 
for the ionization energy for $H^-$ at $B = 100.0\;a.u.$ shows an improvement of more 
than one \% compared to the best value given in the literature. 

For the $^3\Sigma_g$ state we encounter a monotonous decrease in the total energy 
which is proportional to the field strength, and which arises due to the 
spin-Zeeman shift in the presence of a magnetic field. 
In the range $0.0 \le B \le 0.5\;a.u.$ the dissociation energy decreases approximately 
by a factor of two compared to the dissociation energy in field-free space. Simultaneously the equilibrium 
internuclear distance decreases slightly from $1.87$ to $1.81\;a.u.$. 
With further increasing magnetic-field strength we observe a drastical increase in 
the dissociation energy and a simultaneous decrease of the internuclear quilibrium distance. 
A more detailed investigation of the PEC for the $^3\Sigma_g$ state 
shows a second minimum for magnetic-field strengths between $0.05$ and $50.0\;a.u.$ 
which has no counterpart in field-free space. However, this additional minimum is very shallow. 
Vibrational levels were found to exist within the first well of the PEC for the entire range 
of field strengths from $B=0$ to $100\;a.u.$. For the second minimum for field strengths 
$B\gtrsim5\;a.u.$ the existence of vibrational levels depends on the detailed dynamical behavior 
and cannot be decided within the present approach.  
For $B\lesssim5\;a.u.$ a few vibrational levels exist. 

Finally we investigated the lowest $^3\Sigma_u$ state of the \wa which is known to be 
repulsive in field-free space and possesses only a very shallow van der Waals minimum at 
$R \sim 8.0\;a.u.$. The repulsive character of the PEC of the $^3\Sigma_u$ state 
remains for arbitrary field strengths up to $B = 100.0\;a.u.$. 
Due to the spin-Zeeman shift in a magnetic field a crossing occurs between the $^1\Sigma_g$ 
and $^3\Sigma_u$ state in the range $0.1 \le B \le 0.2\;a.u.$. Therefore the lowest state 
of the $\Sigma$ subspace of the \wa in the presence of a magnetic field is the 
$^1\Sigma_g$ state for $B \le 0.1\;a.u.$ and the $^3\Sigma_u$ state for $B \ge 0.2\;a.u.$. 
In superstrong magnetic fields $B \gtrsim 3\times10^3\;a.u.$ the ground state is the $^3\Pi_u$ state. 
The determination of the crossing of the $^3\Sigma_u$ and $^3\Pi_u$ states is a task 
which is left to a future investigation. Furthermore the nonexistence of a strongly bound ground state 
of the \wa has to be confirmed by investigations concerning arbitrary angles between the internuclear 
axis and magnetic-field axis. 
After considering these general properties of the $^3\Sigma_u$ state we investigated the 
development of the van der Waals minimum depending on the field strength. 
For the van der Waals minimum we observe a monotonous decrease in the equilibrium internuclear 
distance with increasing field strength. First the dissociation energy 
increases with increasing field strength and for $B \ge 0.5\;a.u.$ the dissociation energy drastically 
decreases. For $B \ge 1.0\;a.u.$ no minimum has been found. 
No vibrational levels exist for any field strength up to $1\;a.u.$. 

The existence of a minimum which supports a vibrational frequency for the lowest electronic state of
the \wa for intermediate magnetic-field strengths is of particular interest to astrophysics 
in order to determine whether hydrogen molecules exist in the vicinity of white dwarfs. 
According to our investigations for $B \gtrsim 0.2\;a.u.$ the ground state of the \wa 
is not strongly bound and exhibits only a weak minimum due to the van der Waals interaction. 
However, in superstrong magnetic fields the ground state is strongly bound, and the crossing 
between these two states is not known so far. 
Finally we emphasize that for drawing definite conclusions about the existence or nonexistence
of a potential well, the case of nonparallel internuclear axis and magnetic-field axis 
has to be investigated. 
The determination of the corresponding potential energy surfaces is a complicated task which 
is left to future investigations.

\section{Acknowledgments}

The Deutsche Forschungsgemeinschaft as well as the European Community (F. K. D.) 
are gratefully acknowledged for financial support. We thank U. Kappes for fruitful discussions.
Computer time was generously provided by the Rechenzentrum Heidelberg, and in particular 
by the Rechenzentrum Karlsruhe.

\begin{appendix}
\section{construction of the molecular electronic wave function}
In order to construct the spatial part of the electronic wave function for the \wa 
we first establish the molecular orbitals $\Phi_i$ 
for the $H_{2}^+$ ion. The construction of these orbitals was
described in detail in Ref. \cite{kappes:1995}. 
The molecular electronic wave functions for the $H_{2}^+$-ion read as follows:
\begin{eqnarray}
\psi^{\pm m,p} &=& \sum\limits_k c_k
\Phi_k^{\pm m,p}\left(\bbox{r},\alpha_k,\beta_k,R\right) \nonumber \\
&=& \sum\limits_k c_k 
\sum\limits_a^{|m|} \sum\limits_b^N \left( \begin{array}{c}
|m| \\ a \end{array} \right) \left( \begin{array}{c} N \\ b \end{array} \right)
%|m| \\[-0.3 cm] a \end{array} \right) \left( \begin{array}{c} N \\[-0.3 cm] b \end{array} \right)
%\left(\pm i\right)^a \times \nonumber \\
%& & \hspace{0.6 cm} \left [ \phi_{\bbox{n}_{ab,k}} \left( \bbox{r},
\left(\pm i\right)^a \left [ \phi_{\bbox{n}_{ab,k}} \left( \bbox{r},
\alpha_k,\beta_k,+ R/2 \right) + \left(-1\right)^{P + P_{a,k}} \phi_{\bbox{n}_{ab,k}} \left( \bbox{r},
\alpha_k,\beta_k,- R/2 \right) \right] \label{a1}
\end{eqnarray}
Each molecular orbital $\Phi_i$ is an eigenfunction to the molecular angular
momentum operator $L_z$ and to the molecular parity operator $P$ of the $H_2^+$-ion.
Therefore, the $H_2^+$ orbitals $\Phi_i$ are labeled by the eigenvalues $\pm m$ of $L_z$ 
and the eigenvalues $p = \pm 1$ of $P$.
$P_a$ is the parity operator of the atomic orbital and $c_i$ denote the expansion coefficients.
The atomic orbitals $\phi$ building the molecular orbitals $\Phi$
have the following structure:
\begin{equation}
\phi_{\bbox{n}_{ab,k}} \left( \bbox{r},\alpha_k,\beta_k,\pm R/2 \right) = 
x^{n_{x,ab,k}} y^{n_{y,ab,k}} \left(z\mp R/2\right)^{n_{z,ab,k}} exp\left\{- \alpha_k 
\left(x^2 + y^2\right) - \beta_k\left( z\mp R/2 \right)^2 \right\} \label{a2} ,
\end{equation}
where $\alpha_k$ and $\beta_k$ are optimized variational parameters \cite{schmelcher3:1988}.
$R$ represents the internuclear distance and $\bbox{r}^T = (x,y,z)$ the electronic position vector.
For the dependencies of the powers $n_x,n_y,$ and $n_z$ on the atomic magnetic quantum number,
we refer the reader to Ref. \cite{kappes:1995}.
Now we are able to describe the molecular electronic wave function by a linear combination
of products of these $H_{2}^+$ orbitals.
First, we have to ensure that the $H_2$ molecular orbitals are eigenfunctions to the 
molecular angular momentum operator $L_z$ of $H_2$.
Therefore, we arrive at the condition $M = m_1 + m_2$ for the magnetic quantum numbers 
$m_1$ and $m_2$ characterizing the two $H_2^+$ orbitals. 
Moreover, the $H_2$ molecular orbitals have to be eigenfunctions to the 
molecular parity operator $P_g$ of the $H_2$ molecule and the total molecular electronic wave function 
has to respect the Pauli principle.  
Therefore, the electronic wave function of the \wa reads as follows:
\begin{eqnarray}
\Psi^{M,P_g} &=& \sum\limits_{i,j} c_{ij} \left\{ \label{a3}
\Phi_i^{\pm m_1,+1}\left(\bbox{r_1},\alpha_i,\beta_i,R\right)
\Phi_j^{\pm m_2,+1}\left(\bbox{r_2},\alpha_j,\beta_j,R\right)
\pm \Phi_i^{\pm m_1,+1}\left(\bbox{r_2},\alpha_i,\beta_i,R\right)
\Phi_j^{\pm m_2,+1}\left(\bbox{r_1},\alpha_j,\beta_j,R\right) \right\} \nonumber \\
& & \hspace{0.3cm} + c_{ij}' \left\{
\Phi_i^{\pm m_1,-1}\left(\bbox{r_1},\alpha_i,\beta_i,R\right)
\Phi_j^{\pm m_2,-1}\left(\bbox{r_2},\alpha_j,\beta_j,R\right)
\pm \Phi_i^{\pm m_1,-1}\left(\bbox{r_2},\alpha_i,\beta_i,R\right)
\Phi_j^{\pm m_2,-1}\left(\bbox{r_1},\alpha_j,\beta_j,R\right) \right\}
%\Psi^{M,P_g} &=& \sum\limits_{i,j} c_{ij} \left\{ \label{a3}
%\Phi_i^{\pm m_1,+1}\left(\bbox{r_1},\alpha_i,\beta_i,R\right)
%\Phi_j^{\pm m_2,+1}\left(\bbox{r_2},\alpha_i,\beta_i,R\right) \right.\\
%& & \hspace{1.5 cm} \left. \pm \Phi_i^{\pm m_1,+1}\left(\bbox{r_2},\alpha_i,\beta_i,R\right)
%\Phi_j^{\pm m_2,+1}\left(\bbox{r_1},\alpha_i,\beta_i,R\right) \right\} \nonumber \\
%& & \hspace{0.3cm} + c_{ij}' \left\{
%\Phi_i^{\pm m_1,-1}\left(\bbox{r_1},\alpha_i,\beta_i,R\right)
%\Phi_j^{\pm m_2,-1}\left(\bbox{r_2},\alpha_i,\beta_i,R\right) \right. \nonumber \\
%& & \hspace{1.5 cm} \left. \pm \Phi_i^{\pm m_1,-1}\left(\bbox{r_2},\alpha_i,\beta_i,R\right)
%\Phi_j^{\pm m_2,-1}\left(\bbox{r_1},\alpha_i,\beta_i,R\right) \right\} \nonumber
\end{eqnarray}
with 
\begin{eqnarray}
\Phi_i^{\pm m_1,\pm 1}\left(\bbox{r}_{1,2},\alpha_i,\beta_i,R\right) &=& 
\sum\limits_i c_i \sum\limits_a^{|m_1|} \sum\limits_b^N \left( \begin{array}{c}
%|m_1| \\[-0.3 cm] a \end{array} \right) \left( \begin{array}{c} N \\[-0.3 cm] b \end{array} \right)
|m_1| \\ a \end{array} \right) \left( \begin{array}{c} N \\ b \end{array} \right)
\left(\pm i\right)^a \times \nonumber \\
& & \left [ \phi_{\bbox{n}_{ab,i}} \left( \bbox{r}_{1,2},
\alpha_i,\beta_i,+ R/2 \right) + \left(-1\right)^{P_1+P_{a,i}} 
\phi_{\bbox{n}_{ab,i}} \left( \bbox{r}_{1,2},
\alpha_i,\beta_i,- R/2 \right) \right] \nonumber \\ \label{a4}
\end{eqnarray}
and 
\begin{eqnarray}
\Phi_j^{\pm m_2,\pm 1}\left(\bbox{r}_{1,2},\alpha_j,\beta_j,R\right) &=& 
\sum\limits_j c_j \sum\limits_a^{|m_2|} \sum\limits_b^{N'} \left( \begin{array}{c}
%|m_2| \\[-0.3 cm] a \end{array} \right) \left( \begin{array}{c} {N'} \\[-0.3 cm] b \end{array} \right)
|m_2| \\ a \end{array} \right) \left( \begin{array}{c} {N'} \\ b \end{array} \right)
\left(\pm i\right)^a \times \nonumber \\
& & \left [ \phi_{\bbox{n}_{ab,j}} \left( \bbox{r}_{1,2},
\alpha_j,\beta_j,+ R/2 \right) + \left(-1\right)^{P_g + P_2 + P_{a,j}}
\phi_{\bbox{n}_{ab,j}} \left( \bbox{r_{1,2}},
\alpha_j,\beta_j,- R/2 \right) \right] \nonumber \\ \label{a5}
\end{eqnarray}
In order to ensure that the molecular wave function (\ref{a3}) is an eigenfunction 
to the molecular parity operator $P_g$ we multiplied the second atomic orbital 
$\phi_{\bbox{n}_{ab,j}}$ in Eq. (\ref{a5}) by an additional factor of $\left(-1\right)^{P_g}$.
 
In Eq. (\ref{a3}) we have to pay attention to the special case 
when two atomic orbitals belonging to the molecular orbitals $\Phi_i$, Eq. (\ref{a4}), 
and $\Phi_j$, Eq. (\ref{a5}), are equal. 
For both wave functions with ungerade symmetry ($^1\Sigma_u$ and $^3\Sigma_u$)
only one combination of $H_2^+$ orbitals in eq.(\ref{a3}) contributes to the total wave function. 
In that case the two expressions in parenthesis are linear dependent, i.e., they are equal 
except for a multiplication factor 
of $+1$ or $-1$ for the $^1\Sigma_u$ and $^3\Sigma_u$ states, respectively. 
For the $^3\Sigma_g$ state we have no contribution to the total wave function since 
the two expressions in parentheses both equal zero due to the Pauli principle.

\section{evaluation of the matrix elements}
Let us introduce the abbreviation:
\begin{equation}
\Phi_i\left(\bbox{r}\right) := \Phi_i^{\pm m_1,\pm 1}\left(\bbox{r},\alpha_i,\beta_i,R\right)
\end{equation}
with the $H_2^+$ orbitals $\Phi_i$ given in Appendix A. The following 
matrix elements have to be evaluated:
\begin{equation}
\int d\bbox{r}_1\Phi_i \left(\bbox{r}_1\right) \Phi_k \left(\bbox{r}_1\right) \label{b1}
\end{equation}
\begin{equation}
\int  d\bbox{r}_1\Phi_i \left(\bbox{r}_1\right) \label{b2}
\left\{\bbox{p}_1 - \frac{1}{2}\left( \bbox{B}\times \bbox{r}_1\right) \right\}^2 
\Phi_k \left(\bbox{r}_1\right)
\end{equation}
\begin{equation}
\int d\bbox{r}_1 \Phi_i \left(\bbox{r}_1\right) \frac{1}{|\bbox{r}_1 \pm R/2|} \label{b3}
\Phi_k \left(\bbox{r}_1\right) 
\end{equation}
\begin{equation}
\int d\bbox{r}_1 d\bbox{r}_2\Phi_i \left(\bbox{r}_1\right)  \Phi_j \left(\bbox{r}_2\right)  \label{b4}
\frac{1}{|\bbox{r}_1 - \bbox{r}_2|}
\Phi_k \left(\bbox{r}_1\right) \Phi_l \left(\bbox{r}_2\right)
\end{equation}
All these integrals are evaluated in Cartesian coordinates.
For the simple overlap matrix elements in eq.(\ref{b1}) a closed-form  
analytical expression can be given which has been implemented and 
optimized with respect to the numerical performance.
The matrix elements for the kinetic, paramagnetic, and diamagnetic operator in Eq. (\ref{b2}) can be
reduced to simple overlap matrix elements eq.(\ref{b1}).

The evaluation of the electron-nucleus (Eq. \ref{b3}) and electron-electron (Eq. \ref{b4}) 
integrals is much more complicated.
First, one has to regularize the singularities. This is done by 
the following transformation at the expense of an additional integration \cite{singer:1960}:
\begin{eqnarray}
\frac{1}{f\left(\bbox{r}\right)} & = & \pi^{-1/2} \int_{-\infty}^{+\infty} 
exp\left[-u^2 f^2\left(\bbox{r}\right)\right] du \label{b5} \nonumber \\
& = & \frac{2}{\pi^{1/2}} \int_0^1 exp\left[-f^2\left(\bbox{r}\right) \frac{v^2}{1-v^2}\right]
\frac{dv}{\left(1-v^2\right)^{3/2}}
\end{eqnarray}
In our case, $f\left(\bbox{r}\right)$ equals ${|\bbox{r}_1 \pm R/2|}$ 
for the electron-nucleus integral or $|\bbox{r}_1 - \bbox{r}_2|$ 
for the electron-electron integral, respectively.

In the case of the electron-nucleus integral, we first perform the integration
over the electronic coordinates. This integration is done by a special kind of exact quadrature, 
the so-called Rys quadrature \cite{szego:1959,king:1976}. 
The Rys quadrature has been proved to be very useful for the fast calculation 
of two electron integrals, in particular for higher angular momenta of the 
involved orbitals \cite{dupuis:1965,dupuis:1976}.
We emphasize that this quadrature technique yields the exact result for the integration.
Only one integration over the $u$ coordinate remains to be done.
This last integration is performed by a numerical algorithm.
We herefore used a modified Clenshaw-Curtis quadrature based on a CERN library
called 'CHEBQU'. In order to enhance the performance of our calculations, 
the algorithm of this routine
was rewritten in the following way: Normally the routine is intended to perform
one integration at each step. The molecular orbitals $\Phi_i$ in Eqs. 
(\ref{b1}), (\ref{b2}), (\ref{b3}), (\ref{b4}) 
are constructed by means of different atomic orbitals $\phi_n$ in Cartesian coordinates (\ref{a2}).
Within these integrals, only the powers of the polynomials in the electronic $x$ and $y$ coordinates change
but the power of the $z$ coordinate remains the same. It is therefore useful to perform all
integrals over Cartesian coordinates belonging to a molecular orbital $\Phi_i$ in one step.
We hereby avoid the repeated numerical integration of identical integrals.
For the special case of sufficiently small powers of the z-coordinate we implemented explicitly
the analytical solutions of the integration over the $z$ coordinate, i.e., 
we implemented 60 different functions for the different z-integrations.
For a matrix with a dimension of 3500 we are able to evaluate all integrals 
of the forms Eqs. (\ref{b1}),(\ref{b2}), (\ref{b3}) within six seconds 
of CPU time on an AIX 590 computer. 

The electron-electron integrals are the most difficult ones to solve.
Again we first remove the singularity of the integrand by transformation (\ref{b5}).
Subsequently six integrations over the electronic coordinates are performed with the Rys quadrature 
technique.
In total this corresponds to a rather lengthy calculation.
Finally the last integration is again done with a modified Clenshaw-Curtis quadrature. 
Following these steps for the evaluation of the electron-electron integrals
leads to serious trouble. Integrals centered at different nuclei  
turn out to possess very similar values. In fact, many numerically evaluated 
integrals are identical within the accuracy of our numerical integration. 
Therefore, if we simply add the results of different integrals, the desirable 
accuracy is easily lost. To avoid this problem the corresponding integrands 
have been combined is a suitable manner: Instead of adding the results of different integrals 
we added the integrands and then numerically performed the integration. 
We hereby got rid of the accuracy problems. 
Due to the new kind of integrands the termination condition 
of the standard Clenshaw-Curtis quadrature 
proved to be no longer valid. We therefore implemented a new criterion 
to ensure the convergence of our numerical integration.
Similar to the electron-nucleus integrals we used a variety of different functions 
for the special cases of the second $z$ integration. 
The size of the program package for the evaluation of the electron-electron 
and electron-nucleus integrals (programming language C) amounts to 12000 lines.
\end{appendix}

\newpage

\begin{table}
\caption{Data for the lowest $^1\Sigma_g$ state: 
Total $E_{t1}$ and dissociation $E_{d}$ 
energies at the equilibrium internuclear distancec, 
the equilibrium internuclear distances $R_{eq}$, the positions $R_{max}$ and total $E_{t2}$ energies 
for the maximum and the total energies in the dissociation limit 
$\lim\limits_{R\to\infty} E_{t}$.
(all quantities are given in atomic units). 
Transition states appear for field strengths larger than $1.0\;a.u.$.} 
\begin{tabular}{ddddddd}
\multicolumn{1}{c}{\rule[-5mm]{0mm}{11mm}{\raisebox{-0.5ex}[0.5ex]{B}}} & 
\multicolumn{1}{c}{\raisebox{-0.5ex}[0.5ex]{$R_{eq}$}} &
\multicolumn{1}{c}{\raisebox{-0.5ex}[0.5ex]{$E_{d}$}} &
\multicolumn{1}{c}{\raisebox{-0.5ex}[0.5ex]{$E_{t1}$}} &
\multicolumn{1}{c}{\raisebox{-0.5ex}[0.5ex]{$R_{max}$}} &
\multicolumn{1}{c}{\raisebox{-0.5ex}[0.5ex]{$E_{t2}$}} &
\multicolumn{1}{c}{$\lim\limits_{R\to\infty} E_{t}$} \\ \hline
0.0   & 1.40  & 0.173438 & $-$1.173436  & \insmin  & \zinsmin     & $-$0.999998  \\
      &       & 0.173892 & $-$1.173892 \tablenote{larger basis set (see text)}  & & \\
0.001 & 1.40  & 0.173438 & $-$1.173436  & \insmin  & \zinsmin     & $-$0.999998  \\
0.005 & 1.40  & 0.173440 & $-$1.173424  & \insmin  & \zinsmin     & $-$0.999984  \\
0.01  & 1.40  & 0.173450 & $-$1.173396  & \insmin  & \zinsmin     & $-$0.999946  \\
0.05  & 1.40  & 0.173658 & $-$1.172407  & \insmin  & \zinsmin     & $-$0.998750  \\
0.1   & 1.39  & 0.174608 & $-$1.169652  & \insmin  & \zinsmin     & $-$0.995043  \\
0.2   & 1.39  & 0.178001 & $-$1.158766  & \insmin  & \zinsmin     & $-$0.980756  \\
0.5   & 1.33  & 0.194663 & $-$1.089082  & \insmin  & \zinsmin     & $-$0.894419  \\
1.0   & 1.24  & 0.228031 & $-$0.890336  & 11.70    & $-$0.662307  & $-$0.662305  \\
2.0   & 1.09  & 0.291170 & $-$0.335574  &  7.02    & $-$0.044395  & $-$0.044405  \\
5.0   & 0.86  & 0.438015 &    1.801212  &  5.15    &    2.239295  &    2.239227  \\
10.0  & 0.70  & 0.615473 &    5.889023  &  4.23    &    6.504650  &    6.504496  \\
50.0  & 0.417 & 1.371618 &   42.592815  &  2.86    &  43.964915   &   43.964433  \\
100.0 & 0.334 & 1.913452 &   90.506974  &  2.48    &  92.421063   &   92.420426  \\
\end{tabular}
\label{table1}
\end{table}

\begin{table}
\caption{Comparison of the total energies $E_{t}$ as well as 
equilibrium internuclear distances $R_{eq}$ for the lowest $^1\Sigma_g$ state 
with existing results in the literature in the presence of a magnetic field
(all quantities are given in atomic units except for the last column which is in per cent)}
\begin{tabular}{dddddddd}
&\multicolumn{2}{c}{Ref. \cite{turbiner:1983}} &
\multicolumn{2}{c}{Ref. \cite{basile:1987} \tablenote{Energy and equilibrium distance 
may vary to some extent since only PECs and no numerical data 
are given in Ref. \cite{basile:1987}.}} &
\multicolumn{2}{c}{Present work} \\
\cline{2-3} \cline{4-5} \cline{6-7}
\multicolumn{1}{c}{\rule[-5mm]{0mm}{11mm}{\raisebox{-0.5ex}[0.5ex]{\hspace{1.0cm}B}}} & 
\multicolumn{1}{c}{\raisebox{-0.5ex}[0.5ex]{$R_{eq}$}} &
\multicolumn{1}{c}{\raisebox{-0.5ex}[0.5ex]{$E_{t}$}} &
\multicolumn{1}{c}{\raisebox{-0.5ex}[0.5ex]{$R_{eq}$}} &
\multicolumn{1}{c}{\raisebox{-0.5ex}[0.5ex]{$E_{t}$}} &
\multicolumn{1}{c}{\raisebox{-0.5ex}[0.5ex]{$R_{eq}$}} &
\multicolumn{1}{c}{\raisebox{-0.5ex}[0.5ex]{$E_{t}$}} & 
\multicolumn{1}{c}{\raisebox{-0.5ex}[0.5ex]{$\Delta_E\;\left[\%\right]$ 
\tablenote{$\Delta$ is the improvement in the total energy of our results on the existing data 
of Refs. \cite{turbiner:1983,basile:1987}, respectively.}}} \\ \hline
0.4254414 & 1.337 & $-$1.0822  &      &      &  1.349  & $-$1.110362&  2.60\\
0.5       &       &            & 1.29 & 1.06 &  1.33   & $-$1.089082&  2.74 \\
1.0       &       &            & 1.17 & 0.83 &  1.24   & $-$0.890336&  7.27 \\
2.127207  & 1.203 & $-$0.1811  &      &      &  1.070  & $-$0.255591&  41.13\\
4.254414  & 0.859 &    1.3326  &      &      &  0.898  &    1.233808&  7.41\\
21.27207  & 0.528 &   16.0595  &      &      &  0.550  &   15.849134&  1.33\\
42.54414  & 0.463 &   35.76    &      &      &  0.440  &   35.558125&  0.57\\
\end{tabular}
\label{table2}
\end{table}

\begin{table}
\caption{Data for the lowest $^1\Sigma_u$ state: Total $E_{t}$ and 
dissociation $E_{d}$ energies at the equilibrium internuclear distance, 
the equilibrium internuclear distances $R_{eq}$ and total energies in the dissociation limit 
$\lim\limits_{R\to\infty} E_{t}$ 
(all quantities are given in atomic units)} 
\begin{tabular}{ddddd}
\multicolumn{1}{c}{\rule[-5mm]{0mm}{11mm}{\raisebox{-0.5ex}[0.5ex]{B}}} & 
\multicolumn{1}{c}{\raisebox{-0.5ex}[0.5ex]{$R_{eq}$}} &
\multicolumn{1}{c}{\raisebox{-0.5ex}[0.5ex]{$E_{d}$}} &
\multicolumn{1}{c}{\raisebox{-0.5ex}[0.5ex]{$E_{t}$}} &
\multicolumn{1}{c}{$\lim\limits_{R\to\infty} E_{t}$} \\ \hline
0.0      &  2.42  & 0.131112 &  $-$0.756111 & $-$0.624999  \\
0.001    &  2.42  & 0.131112 &  $-$0.756109 & $-$0.624997  \\
0.005    &  2.42  & 0.131140 &  $-$0.756097 & $-$0.624957  \\
0.01     &  2.42  & 0.131246 &  $-$0.756069 & $-$0.624823  \\
0.05     &  2.43  & 0.132568 &  $-$0.753408 & $-$0.620839  \\
0.1      &  2.46  & 0.136370 &  $-$0.746347 & $-$0.609977  \\
0.2      &  2.53  & 0.147250 &  $-$0.722812 & $-$0.575562  \\
0.5      &  2.50  & 0.180626 &  $-$0.602596 & $-$0.421969  \\
1.0      &  2.30  & 0.225593 &  $-$0.316748 & $-$0.091155  \\
2.0      &  2.01  & 0.291037 &   0.389051 &  0.680087  \\
5.0      &  1.60  & 0.414834 &   2.857162 &  3.271996  \\
10.0     &  1.30  & 0.542594 &   7.327008 &  7.869602  \\
50.0     &  0.666 & 0.837018 &  45.524983 & 46.362001  \\
100.0    &  0.490 & 1.046760 &  94.391271 & 95.438032  \\
\end{tabular}
\label{table3}
\end{table}

%%%%%%%%%%% IMPORTANT !!!!!!!!!!!!!!!!!!!!
% correct table is located in ~thomasd/magnet/data/h2/table/ !!!!!
%%%%%%%%%%%%%%%

\begin{table}
\squeezetable
\caption{Total $E_{t1},E_{t2}$ and dissociation energies $E_{d1},E_{d2}$ at the 
equilibrium internuclear distance, 
equilibrium internuclear distances $R_{eq1},R_{eq2}$ and total energies in the dissociation limit 
$\lim\limits_{R\to\infty} E_{t}$ for the lowest $^3\Sigma_g$ state 
(all quantities are given in atomic units)} 
\begin{tabular}{dddddddd}
\multicolumn{1}{c}{\rule[-5mm]{0mm}{11mm}{\raisebox{-0.5ex}[0.5ex]{B}}} & 
\multicolumn{1}{c}{\raisebox{-0.5ex}[0.5ex]{$R_{eq1}$}} &
\multicolumn{1}{c}{\raisebox{-0.5ex}[0.5ex]{$E_{d1}$}} &
\multicolumn{1}{c}{\raisebox{-0.5ex}[0.5ex]{$E_{t1}$}} &
\multicolumn{1}{c}{\raisebox{-0.5ex}[0.5ex]{$R_{eq2}$}} &
\multicolumn{1}{c}{\raisebox{-0.5ex}[0.5ex]{$E_{d2}$}} &
\multicolumn{1}{c}{\raisebox{-0.5ex}[0.5ex]{$E_{t2}$}} &
\multicolumn{1}{c}{$\lim\limits_{R\to\infty} E_{t}$} \\ \hline
0.0   &  1.87  & 0.112124 & $-$0.737124  & \insmin& \zinsmin  & \zinsmin    & $-$0.624999  \\
0.001 &  1.87  & 0.112104 & $-$0.738101  & \insmin& \zinsmin  & \zinsmin    & $-$0.625997  \\
0.005 &  1.87  & 0.112071 & $-$0.742027  & \insmin& \zinsmin  & \zinsmin    & $-$0.629957  \\
0.01  &  1.87  & 0.111865 & $-$0.746688  & \insmin& \zinsmin  & \zinsmin    & $-$0.634823  \\
0.05  &  1.86  & 0.108885 & $-$0.779724  & 10.82  & 6.29$\times10^{-4} $& $-$0.671469  & $-$0.670839  \\
0.1   &  1.84  & 0.101538 & $-$0.811515  & 10.55  & 6.90$\times10^{-4} $& $-$0.710667  & $-$0.709977  \\
0.2   &  1.82  & 0.085048 & $-$0.860609  &  9.72  & 1.004$\times10^{-3} $& $-$0.776566  & $-$0.775562  \\
0.5   &  1.81  & 0.062317 & $-$0.984286  &  8.17  & 1.603$\times10^{-3} $& $-$0.923572  & $-$0.921969  \\
1.0   &  1.76  & 0.065670 & $-$1.156825  &  7.18  & 1.910$\times10^{-3} $& $-$1.093065  & $-$1.091155  \\
2.0   &  1.55  & 0.090984 & $-$1.410897  &  6.91  & 1.660$\times10^{-3} $& $-$1.321573  & $-$1.319913  \\
5.0   &  1.20  & 0.168071 & $-$1.896074  &  7.35  & 9.66$\times10^{-4} $& $-$1.728970  & $-$1.728004  \\
10.0  &  0.96  & 0.274453 & $-$2.404851  &  7.68  & 6.13$\times10^{-4} $& $-$2.131011  & $-$2.130398  \\
50.0  &  0.560 & 0.781791 & $-$4.245251  &  8.19  & 2.11$\times10^{-4} $& $-$3.463671  & $-$3.463461  \\
100.0 &  0.446 & 1.162187 & $-$5.415591  &\insmin & \zinsmin  & \zinsmin    & $-$4.253404  \\
\end{tabular}
\label{table4}
\end{table}

%\begin{table}
%\caption{Position $R_{max}$ and total energies $E_{t}$ for the maximum of the lowest $^3\Sigma_g$ state 
%(all quantities are given in atomic units)} 
%\begin{tabular}{ddd}
%\multicolumn{1}{c}{\rule[-5mm]{0mm}{11mm}{\raisebox{-0.5ex}[0.5ex]{B}}} & 
%\multicolumn{1}{c}{\raisebox{-0.5ex}[0.5ex]{$R_{max}$}} &
%\multicolumn{1}{c}{\raisebox{-0.5ex}[0.5ex]{$E_{t}$}} \\ \hline 
%0.05     & 7.95  & $-$0.671180 \\
%0.1      & 6.07  & $-$0.709479 \\
%0.2      & 5.93  & $-$0.774682 \\
%0.5      & 5.41  & $-$0.922264  \\
%1.0      & 5.21  & $-$1.092420 \\
%2.0      & 4.81  & $-$1.320663  \\
%5.0      & 3.99  & $-$1.724693 \\
%10.0     & 3.46  & $-$2.120833 \\
%50.0     & 2.61  & $-$3.430287 \\
%100.0    & 2.36  & $-$4.208860 \\
%\end{tabular}
%\label{table5}
%\end{table}

\begin{table}
\caption{Data for the lowest $^3\Sigma_u$ state: Total $E_{t}$ and 
dissociation $E_{d}$ energies at the equilibrium internuclear distance, 
the equilibrium internuclear distances $R_{eq}$ and total energies in the dissociation limit 
$\lim\limits_{R\to\infty} E_{t}$ 
(all quantities are given in atomic units).} 
\begin{tabular}{ddddd}
\multicolumn{1}{c}{\rule[-5mm]{0mm}{11mm}{\raisebox{-0.5ex}[0.5ex]{B}}} & 
\multicolumn{1}{c}{\raisebox{-0.5ex}[0.5ex]{$R_{eq}$}} &
\multicolumn{1}{c}{\raisebox{-0.5ex}[0.5ex]{$E_{d}$}} &
\multicolumn{1}{c}{\raisebox{-0.5ex}[0.5ex]{$E_{t}$}} &
\multicolumn{1}{c}{$\lim\limits_{R\to\infty} E_{t}$} \\ \hline
0.0      &  7.9  & 1.$885\times10^{-5}$& $-$1.0000173 &  0.9999985 \\
0.001    &  7.9  & 1.$897\times10^{-5}$& $-$1.0010170 & $-$1.0009981 \\
0.005    &  7.9  & 1.$913\times10^{-5}$& $-$1.0050030 & $-$1.0049839 \\
0.01     &  7.9  & 1.$919\times10^{-5}$& $-$1.0099656 & $-$1.0099464 \\
0.05     &  7.9  & 1.$917\times10^{-5}$& $-$1.0487690 & $-$1.0487499 \\
0.1      &  7.8  & 1.$935\times10^{-5}$& $-$1.0950628 & $-$1.0950434 \\
0.2      &  7.7  & 2.$060\times10^{-5}$& $-$1.1807764 & $-$1.1807558 \\
0.5      &  7.4  & 1.$532\times10^{-5}$& $-$1.3944338 & $-$1.3944185 \\
1.0      &  7.5  & 0.$373\times10^{-5}$& $-$1.6623090 & $-$1.6623053 \\
2.0      & \insmin& \zinsmin & \zinsmin    & $-$2.0444046 \\
5.0      & \insmin& \zinsmin & \zinsmin    & $-$2.7607732 \\
10.0     & \insmin& \zinsmin & \zinsmin    & $-$3.4955044 \\
50.0     & \insmin& \zinsmin & \zinsmin    & $-$6.0355668 \\
100.0    & \insmin& \zinsmin & \zinsmin    & $-$7.5795742 \\
\end{tabular}
\label{table6}
\end{table}

\end{document}